\documentclass[prd,floats,superscriptaddress,nofootinbib,preprintnumbers]{revtex4}

\usepackage[normalem]{ulem}

\usepackage{lipsum, babel}

\usepackage[a4paper, margin=2cm]{geometry}

\usepackage{amssymb}
\usepackage{amsmath}

\usepackage{sans}

\usepackage{graphicx}
\usepackage{graphics}
\usepackage{color}
\usepackage{rotate}
\usepackage{fancyhdr} 
\usepackage[hidelinks]{hyperref}
\usepackage{indentfirst}
\usepackage{booktabs} 
\usepackage{fontawesome} 
\usepackage{physics} 
\usepackage{siunitx} 
\DeclareSIUnit \parsec {pc}
\DeclareSIUnit \year {yr}
\usepackage{umoline}
\usepackage{tikz}
\usetikzlibrary{decorations.pathmorphing}  
\usepackage[caption=false]{subfig}

\newcommand{\iu}{\mathrm{i}}
\newcommand{\e}{\mathrm{e}}

\begin{document}
\noindent \makebox[15cm][l]{\footnotesize \hspace*{-2cm} }{\footnotesize 
DESY-25-065}  \\  [-2.8mm]
~\vspace{0cm}
\title{Bracketing the soliton-halo relation of ultralight dark matter}

\author{Kfir Blum} 
\affiliation{Department of Particle Physics and Astrophysics, Weizmann Institute of Science,
Herzl St 234, Rehovot 761001, Israel}
\author{Marco Gorghetto}
\affiliation{Deutsches Elektronen-Synchrotron DESY, Notkestr. 85, 22607 Hamburg, Germany}
\author{Edward Hardy}
\affiliation{Rudolf Peierls Centre for Theoretical Physics, University of Oxford, Parks Road, Oxford OX1 3PU, UK}
 \author{Luca Teodori}\email{luca.teodori@weizmann.ac.il (corresponding author)}  
 \affiliation{Weizmann Institute of Science, Rehovot 7610001, Israel}

\begin{abstract}

In theories of ultralight dark matter, solitons form in the inner regions of galactic halos. The observational implications of these depend on the soliton mass. Various relations between the mass of the soliton and properties of the halo have been proposed. We analyze the implications of these relations, and test them with a suite of numerical simulations. The relation of Schive et al.~2014 is equivalent to $(E/M)_{\rm sol}=(E/M)_{\rm halo}$ where $E_{\rm sol (halo)}$ and $M_{\rm sol (halo)}$ are the energy and mass of the soliton (halo). If the halo is approximately virialized, this relation is parametrically similar to the evaporation/growth threshold of Chan et al.~2022, and it thus gives a rough lower bound on the soliton mass. A different relation has been proposed by Mocz et al.~2017, which is equivalent to $E_{\rm sol}=E_{\rm halo}$, so is an upper bound on the soliton mass provided the halo energy can be estimated reliably. Our simulations provide evidence for this picture, and are in broad consistency with the literature, in particular after accounting for ambiguities in the definition of $E_{\rm halo}$ at finite volume.

\end{abstract}

\maketitle

\tableofcontents
\newpage
\section{Introduction} \label{s:intro}

Ultra-Light Dark Matter (ULDM) is a well-motivated Dark Matter (DM) candidate, potentially arising in high energy completions of the Standard Model of Particle Physics. It is generically produced in the early Universe via the vacuum misalignment mechanism, and is stable on cosmological timescales~\cite{Preskill:1982cy,Abbott:1982af,Dine:1982ah,Svrcek:2006yi,Arvanitaki:2009fg,Marsh:2015xka,Hui2017}. 
Compared to collision-less Cold Dark Matter, ULDM leads to novel behavior on distances comparable to or smaller than its de-Broglie wavelength $\lambda_{\rm dB} = 2\pi/(mv) $, where $ v $ is the characteristic 
velocity of a system. On such scales ULDM's wave-like nature is manifest. 
This results in a suppression of power in cosmological perturbations, leaving observable imprints on the Cosmic Microwave Background (CMB) anisotropy power spectrum, galaxy clustering~\cite{Lague:2021frh}, and the Lyman-alpha forest~\cite{Irsic:2017yje,Armengaud:2017nkf,Kobayashi:2017jcf,Leong:2018opi}. 
ULDM wave-like density fluctuations 
can also lead to astrophysical effects inside galaxies, such as dynamical heating and dynamical friction~\cite{Hui2017,Lancaster:2019mde,DuttaChowdhury:2023qxg}, leading to constraints using systems like dwarf and ultra-faint dwarf galaxies~\cite{Zimmermann:2024xvd,Teodori:2025rul,Dalal:2022rmp}. Observational constraints on the magnitude of such effects bounds the particle mass of an ULDM candidate $m$ that comprises all of Dark Matter to satisfy $ m \gtrsim \SI{e-21}{\electronvolt}$. See e.g.~\cite{Hui2017,Hui:2021tkt,Ferreira:2020fam} 
for reviews.

Another key feature of ULDM, on which we focus in this work, is the formation of cored density profiles at the centers of galaxy halos. These cores consist of `solitons', which are a ground state of the system in the sense that the soliton solution to the ULDM equations of motion minimizes the energy for a fixed mass. Solitons have been seen in ULDM halos in many numerical simulations~\cite{Guzman2004,Schive:2014dra,Schive:2014hza,Schwabe:2016rze,Veltmaat:2016rxo,Mocz:2017wlg,Veltmaat:2018dfz,Levkov:2018kau,Eggemeier:2019jsu,Chen:2020cef,Schwabe:2020eac,Zhang:2018ghp,Schwabe:2021jne}. 
Solitons can affect the observed rotation curves of low-surface-brightness galaxies~\cite{Bar:2018acw,Bar:2021kti} and irregular dwarf galaxies~\cite{Banares-Hernandez:2023axy}, stellar kinematics and dynamics of dwarf galaxies~\cite{Marsh:2015wka,DuttaChowdhury:2023qxg,Teodori:2025rul}, and even strong gravitational lensing time delays~\cite{Blum:2021oxj,Blum:2024igb}, and involve interesting physics such as stochastic motion~\cite{Li:2020ryg,Chowdhury:2021zik} and quasi-normal mode fluctuations~\cite{Guzman2004,Chan:2023crj}\footnote{Other gravitational probes of ULDM include~\cite{Khmelnitsky:2013lxt,Schutz:2020jox,Laroche:2022pjm,Powell:2023jns, Blas:2024duy,Hertzberg:2022vhk}.}. It has been suggested that soliton cores may play a role in resolving the core-cusp problem~\cite{Kendall:2019fep}, namely, the mismatch between simulations of cold dark matter~\cite{Navarro:1996gj} and observations~\cite{Moore:1994yx}.  

A natural question is whether a soliton forms within the lifetime of a galaxy and, if so, what is its expected mass for a given host galaxy halo. 
Dynamical relaxation estimates of the timescale for soliton formation 
are consistent with the results of simulations that use ``noise'' initial conditions, which are designed to be in the kinetic regime~\cite{Bar-Or:2018pxz,Bar-Or:2020tys,Levkov:2018kau,Dmitriev:2023ipv,Chavanis:2020upb}.  
Meanwhile, simulations with cosmological initial conditions -- notably those of Ref.~\cite{Schive:2014dra,May:2021wwp} -- suggest that solitons in cosmological halos may form more rapidly than predicted by 
kinetic theory estimates (the impact of baryons has been studied in Ref.~\cite{Veltmaat:2019hou}). 
Regarding the expected soliton mass, the cosmological simulations in  Ref.~\cite{Schive:2014dra,Schive:2014hza} (see also Ref.~\cite{Liao:2024zkj}) provided numerical evidence for a simple relation between the soliton mass and the host halo mass and energy. 
Those authors also suggested that the relation may represent an attractor of the equations of motion, supporting this point via simulations with different initial conditions. Many other investigations of the soliton-halo relation have subsequently appeared in the literature, reporting varying levels of agreement~\cite{Schive:2014hza,Schwabe:2016rze,Du:2016aik,Mocz:2017wlg,Chan:2021bja,Zagorac:2022xic,Lague:2023wes,Liao:2024zkj,Manita:2024vww,Yavetz:2021pbc}. In this work, we present a new perspective on the problem. A summary of our main understanding is as follows. 

The soliton-halo relation found in Ref.~\cite{Schive:2014hza} was fit in that paper by the expression\footnote{See Eq.~(5) and Fig.~4 {\it there}; we restore natural units to their expression. Ref.~\cite{Schive:2014hza} referred to $ M_{\rm c}$ rather than $ M_{\rm sol} $, where $ M_{\rm c}$ is the soliton mass enclosed in the the core radius $ r_{\rm c} $, at which the density drops to half its maximum value. The relation is $  M_{\rm sol} \approx 4.2 M_{\rm c} $.}
\begin{equation} \label{eq:schive}
    M_{\rm sol} = 4.2 \qty(\frac{|E_{\rm tot}|}{M} )^{1/2} \frac{1}{Gm} \ ,
\end{equation}
where  $M_{\rm sol}$ is the soliton mass. In a cosmological setting $E_{\rm tot}$ and $M$ are the total energy and the virial mass of a halo, including the soliton. In flat-space simulations $E_{\rm tot}$ and $M$ are the total energy and the total mass in the simulation box (both of these are conserved quantities), which are a proxy for the halo properties in a cosmological system, assuming most of the simulated volume is virialized.

The simulations used to obtain Eq.~(\ref{eq:schive}) employed toy initial conditions (multiple solitons starting at rest); however, Ref.~\cite{Schive:2014hza} noted that the result was consistent with cosmological simulations~\cite{Schive:2014dra,Chan:2021bja}. 
The prefactor $\approx4.2$ in Eq.~(\ref{eq:schive}) was obtained in~\cite{Schive:2014hza} by fitting the numerical data.\footnote{Since Ref.~\cite{Schive:2014hza} phrases Eq.~(\ref{eq:schive}) in terms of $M_{\rm c}$,  the prefactor in Eq.~(5) {\it there} is $\alpha=4.2M_{\rm c}/M_{\rm sol}\approx1$.} The exponent $1/2$ on $|E_{\rm tot}|/M$ was also confirmed numerically, with insight from cosmological simulations. 
We will call Eq.~(\ref{eq:schive}) the $1/2$ relation. 

Ref.~\cite{Bar:2018acw} pointed out that the $1/2$ relation is directly equivalent to the relation
\begin{equation}\label{EMEM}
    \frac{E_{\rm sol}}{M_{\rm sol}} = \frac{E_{\rm tot}}{M} \;\;\;\;\;\;{\rm(equivalent\;to\;Eq.\ (1))}\;.
\end{equation}
This follows because the soliton by itself satisfies the relation
$M_{\rm sol} \approx 4.2 \qty(\frac{|E_{\rm sol}|}{M_{\rm sol}} )^{1/2} \frac{1}{Gm},
$ 
as can be 
verified by direct calculation of the soliton solution\footnote{Note that the $1/2$ relation is consistent with the findings of Ref.~\cite{Chavanis:2019faf}.}. We note that it is unlikely that a soliton would have a mass significantly smaller than that determined by the $1/2$ relation. 
This is suggested by the results of Ref.~\cite{Chan:2022bkz}, which investigated the evaporation and growth of a soliton in a homogeneous ``noisy'' background, showing that in such a system solitons with masses smaller by more than a factor of approximately $3$  than those predicted by Eq.~(\ref{EMEM}) would evaporate due to scattering with the background particles. We return to this point in more detail later on.

Another soliton-halo relation was proposed in Ref.~\cite{Mocz:2017wlg}. We call this the $1/3$ relation since it can be rewritten as
\begin{equation} \label{eq:13}
M_{\rm sol} = 2.6 \qty(\frac{|E_{\rm tot}|}{G^2m^2})^{ 1/3 } \ .
\end{equation}
In fact, if the simulated system is approximately virialized, then Eq.(\ref{eq:13}) can only be an upper bound on the soliton mass. This can easily be understood by the fact that, as pointed out in Ref.~\cite{Bar:2018acw},  Eq.~(\ref{eq:13}) is directly equivalent to
\begin{equation} \label{eq:13_simple}
E_{\rm sol} = E_{\rm tot} \ ,
\end{equation}
because an expression similar to Eq.~(\ref{eq:13}) holds for a single soliton:
$M_{\rm sol} \approx 2.64 \qty(\frac{|E_{\rm sol}|}{G^2m^2})^{ 1/3 }$ (see also Appendix~\ref{s:soliton}).  
A soliton more massive than that predicted by the $1/3$ relation would have more energy than the entire halo (in absolute value), which is an impossibility for bound virialized systems. 
The $1/3$ upper bound may be violated if a flat-space simulation contains unbound particles since these would contribute positively to $E_{\rm tot}$; the bound however holds if $E_{\rm tot}$ is restricted to the bound particles.

\begin{figure}
    \centering
\includegraphics[width=0.65 \linewidth]{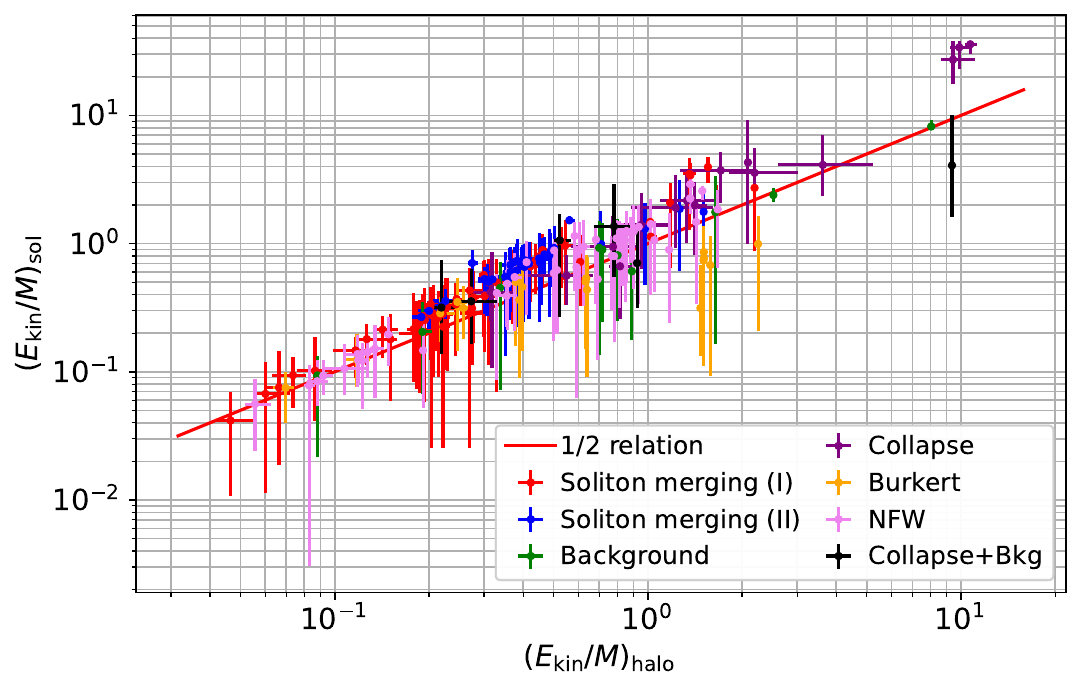}
     \caption{The kinetic energy-to-mass ratio of solitons compared to that of their halos, $(E_{\rm kin}/M)_{\rm sol}$ and $(E_{\rm kin}/M)_{\rm halo}$ respectively. Data points are the results from simulations with all the types of initial conditions that we tested, see Section~\ref{ss:init}. The red line is the $1/2$ relation in Eq.~(\ref{EMEM}), i.e. $(E_{\rm kin}/M)_{\rm sol}=(E_{\rm kin}/M)_{\rm halo}$. After the soliton halo-system forms, both $(E_{\rm kin}/M)_{\rm sol}$ and $(E_{\rm kin}/M)_{\rm halo}$ are time-dependent as the system evolves, with the soliton slowly accreting mass, whereas the halo, after it forms, does not undergo significant evolution (see e.g. bottom-right panel of Fig.~\ref{fig:sol_halo_example}). The error bars on data points represent the change during this evolution as well as the effects of soliton fluctuations, with central values corresponding to the mean values over the evolution (see Fig.~\ref{fig:sol_halo_example}). 
     [The maximum values of $(E_{\rm kin}/M)_{\rm sol}$, $(E_{\rm kin}/M)_{\rm halo}$ are therefore (weakly) sensitive to the simulation run time.] 
     Horizontal error bars correspond to the fluctuations in $E_{\rm kin}$ towards the end of simulations. 
     }
    \label{fig:E_M}
\end{figure}

To test the picture above, we performed simulations of ULDM in flat space-time and studied the soliton-halo configurations arising from the evolution of 
different types of initial conditions, see Section~\ref{s:results} for details. 
The results, collected in Fig.~\ref{fig:E_M}, are consistent with our expectations: the solitons form not far from (and usually below) Eq.~(\ref{EMEM}), and thus from the $1/2$ relation.  
Subsequently, the solitons accrete more mass, but do not  surpass the $1/3$ relation. Fig.~\ref{fig:generic_final_snapshot}, which uses variables that allow both relations to be represented on a single plot, demonstrates this point.

In Section~\ref{s:discussion}, we compare with other works and show that their data is not clearly inconsistent with the band enclosing the $1/2$ and $1/3$ relations (after accounting for an arbitrary additive constant present in the definition of the potential energy for finite box simulations).

If our results can be translated to a cosmological setting (which may very well be the case, see Fig.~\ref{fig:Chan_plot}), ULDM bounds based on the $1/2$ relation may be considered conservative when applied to astrophysical systems, provided that the soliton formation time is not too large compared to the age of the system. 
Assuming the kinetic theory relaxation calculation gives the relevant timescale, the formation of the soliton limits the applicability of the soliton-halo relation in galactic systems 
like dwarf galaxies to ULDM masses $m \lesssim 10^{-20} \, \unit{\electronvolt}$~\cite{Dmitriev:2023ipv}.

The paper is structured as follows. We describe the setup of our simulations in Section~\ref{s:sim}; we present our main results in Section~\ref{s:results}; we discuss implications of our results and comparison with previous works in Section~\ref{s:discussion} and we conclude in Section~\ref{s:sum}. Further details are provided in Appendices: 
In Appendix~\ref{s:code}, we describe the algorithm we use to numerically solve the equations of motion; in Appendix~\ref{s:soliton} we review basics of the soliton solution; in Appendix~\ref{s:initial_cond} we give precise details of the initial conditions used in simulations; in Appendix~\ref{s:sol_form_criterium} we investigate the sensitivity of our results to our choice of criterion for a soliton to have formed. Finally, in Appendix~\ref{s:spectrum} we review how the energy spectrum of the ULDM field can be calculated and discuss additional insights that this can provide. 

\section{Simulation setup
} \label{s:sim}

Due to its huge occupation number, the evolution of ULDM is 
well-approximated by its classical equations of motion (EoM), which, in the non-relativistic limit of interest, reduce to the Schr\"odinger--Poisson equations. We consider systems in boxes with periodic boundary conditions on grids with $256^3$ or $512^3$ points, and evolve them numerically with a standard pseudo-spectral algorithm (see e.g.~\cite{Levkov:2018kau}).  
Key quantities are the energy of a soliton $E_{\rm sol}=E_{\rm kin,sol}+E_{\rm pot,sol}$ and of the halo $E_{\rm tot}=E_{\rm kin}+E_{\rm pot}$, where $E_{\rm kin(,sol)}$ and $E_{\rm pot(,sol)}$  are the kinetic and potential energies. Throughout this work, we discuss systems with zero net angular momentum.
As in all finite-box (non-GR) simulations, the potential energy $E_{\rm pot}$ 
is defined only up to an additive constant, 
and the naive evaluation of this 
is unphysical as it depends on the box size. We choose this constant such that the gravitational potential of the resulting halo matches the behavior that would occur in the infinite-volume limit ($\propto 1/r$), see Appendix~\ref{s:code}.  
Where possible, we present our results using 
$ E_{\rm kin} $ (which is independent of this constant and the uncertainty associated with its determination), 
as we discuss further in Section~\ref{s:results}.

Upon redefining space, time and density as $x\to x/m$, $t\to t/m$, $\rho\to m^2/(4\pi G)\rho$ respectively, the EoM do not depend on $m$. When showing results we however set $m=10^{-21}~{\rm eV}$. Additionally, the EoM retain a residual scaling symmetry (rescaling the field, distances, times and potentials, see Appendix~\ref{s:code}). We use this to fix the  box length to $L=10\,{\rm kpc}$; this choice is useful to connect the DM halos in our simulations 
to observed galactic halos, which have scale radii of $O(10\, \text{kpc})$. In combination with our choice of $m$, this also fixes the units of time and density,  expressed in ${\rm Gyr}$ and $M_\odot/{\rm kpc}^3$ respectively.

\subsection{Initial conditions}\label{ss:init}
We study the soliton-halo configurations arising from  different sets of initial conditions: 
\begin{itemize}
    \item 
    \emph{Soliton merging}.   We start with $N=4-100$ solitons randomly placed and at rest, either with (I) the \emph{same} radius chosen in the range $L/(10-50)$, as in Ref.~\cite{Schive:2014hza}, or  (II) with a \emph{random} radius sampled from a uniform distribution in the range $L/(20-100)$, which resembles  
    the setup of Ref.~\cite{Mocz:2017wlg}.
    \item \emph{Halo}. We initialize the ULDM  field $\psi$ with a density  
    approximating that of a virialized (statistically stationary) Navarro-Frenk-White (NFW)~\cite{Navarro:1996gj} or Burkert~\cite{Salucci:2000ps,Burkert:2015vla} halo profile. The parameters of the halos are chosen to mimic dwarf galaxy systems such as Fornax~\cite{Walker:2008ax,Walker:2009zp} for $ m=\left(10^{-22}-10^{-21}\right)\unit{\electronvolt} $, with initial density parameter $\rho_0= \left(10^{8}-10^{10}\right) \ \unit{M_\odot\per\kilo\parsec\cubed}$ and scale radius $r_{\rm s}= \left(0.5- 2 \right) \unit{\kilo\parsec} $, see Eq.~(\ref{eq:nfw_burk}). 

    \item \emph{Background}. We initialize $ \psi $ as a Gaussian noise  background with mean density $\bar{\rho}$ and Maxwellian velocity distribution, with momentum variance $k_{\rm bkg}^2/2$, as in Refs.~\cite{Levkov:2018kau,Chan:2022bkz}, i.e.
\begin{equation} \label{eq:cov}
	\expval{\psi^*(x) \psi(y)} = C(x-y) \ , \qquad C(x) = \bar{\rho}\,\e^{-x^2k^2_{\rm bkg}/4} \ .
\end{equation}
We consider a sufficiently dense background such that the Jeans length, $\lambda_{\rm J} \equiv 2\pi / (16\pi G \bar{\rho} m^2)^{1/4}$, is smaller than $L$, ensuring that it rapidly forms a gravitationally bound halo via the Jeans instability~\cite{Levkov:2018kau}.
The range of parameters probed is $k_{\rm bkg} L=10-100$ and $\bar{\rho}$ such that $\lambda_{\rm J}/L =0.1- 0.4$. 

\item \emph{Collapse.} We begin with an elliptical overdensity of the form 
\begin{equation}\label{ellipt}
	\psi(x) = \sqrt{\rho_0} \e^{-x_1^2/2R_1^2 -x_2^2/2R_2^2-x_3^2/2R_3^2} \ ,
\end{equation}
with density and radius varying in the range $\rho_0 = \left(0.01-1\right)\ \unit{M_\odot\per\parsec\cubed}$ and $R_i=\left(0.5-5 \right)\ \unit{\kilo\parsec}$. The phase of the field is coherent across the initial configuration, but the overdensity quickly collapses and in the process loses coherence.
\item \emph{Background+Collapse}. We consider the elliptical overdensity in Eq.~(\ref{ellipt}) on top of the noise background in Eq.~(\ref{eq:cov}).
\end{itemize}
Further details can be found in Appendix~\ref{s:initial_cond}.

\subsection{When should a soliton-halo relation be computed?}\label{ss:when}

Our goal is to clarify the properties of systems in which a soliton is embedded in a host halo whose mass is much larger than that of the soliton, $M\gg M_{\text{sol}}$. 
Such hierarchical systems enable kinematic tracer observables (stellar or gas velocities) in the halo's large-scale region to be compared with ULDM's velocity predictions for the inner region, where the soliton lives, enabling the  
soliton to be bounded or potentially discovered~\cite{Bar:2018acw,Teodori:2025rul}.

To this aim, we need to choose a time at which to determine whether a soliton has formed, and, if it has, its properties and soliton-halo relation. In a cosmological context, the obvious phenomenologically-motivated choice is the Hubble time. However, the timescales involved in simulations with some types of initial conditions (e.g., those starting from a background, or soliton mergers) do not necessarily coincide with those of realistic cosmological initial conditions; in other words, the properties of the soliton and halo at a Hubble time in such simulations might not be relevant to cosmology. Nevertheless, we believe that such systems can still elucidate the dynamics leading to physical soliton-halo relation.

We therefore adopt the following procedure to determine a time, $t_{\rm form}$, from which point on we analyze the soliton and halo, with the understanding that this inevitably leads to some arbitrariness: At each timestep in a simulation we identify the point in the box that contains the highest density, $\rho_0$. We then determine whether the spherically-averaged density profile around this point is well-matched by the soliton solution with central density $\rho_0$ and half-density radius $r_{\rm c}(\rho_0)=1.3/(4\pi Gm^2\rho_0)^{1/4}$ for $r<3r_{\rm c}$  and an NFW profile for $r>4r_{\rm c}$ (with scale radius and density as free parameters). If such a fit is sufficiently good, we consider the soliton-halo system as formed. Further details and analysis of the uncertainties arising from our choices can be found in Appendix~\ref{s:sol_form_criterium}, while in Appendix~\ref{s:spectrum} we discuss an alternative way to pinpoint soliton formation, via the spectral energy density. 

\begin{figure}
\qquad  \includegraphics[width=0.4\textwidth]{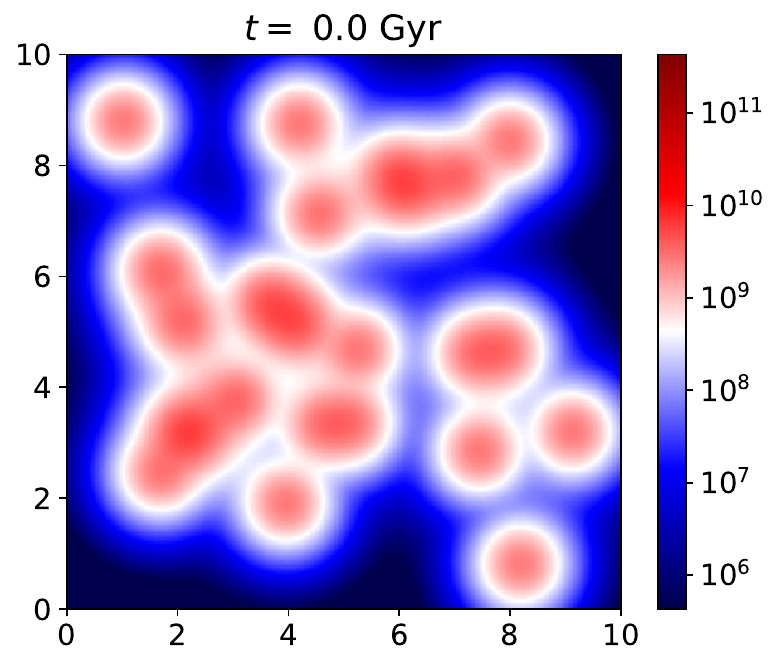} \quad	\includegraphics[width=0.4\textwidth]{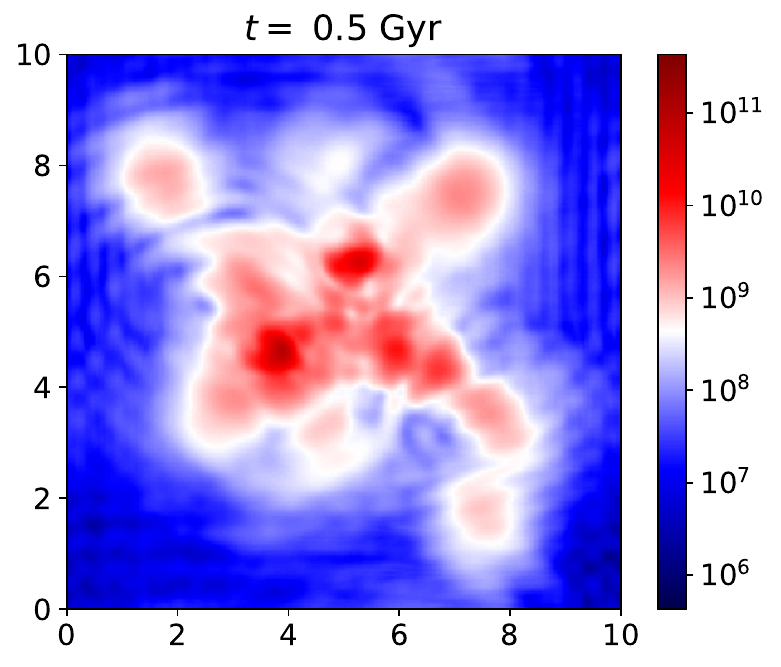} \\
\qquad \includegraphics[width=0.4\textwidth]{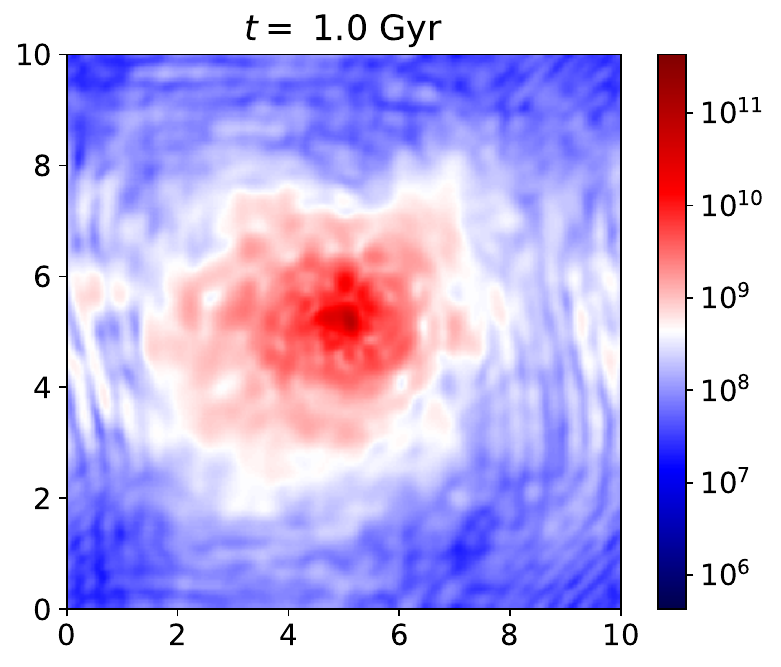} 
\quad
\includegraphics[width=0.4\textwidth]{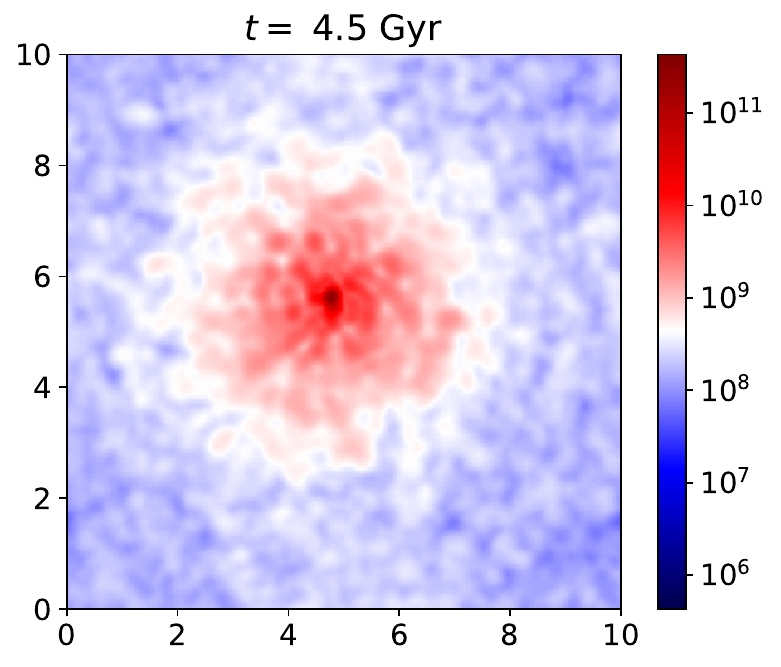}
\includegraphics[width=0.485\textwidth]{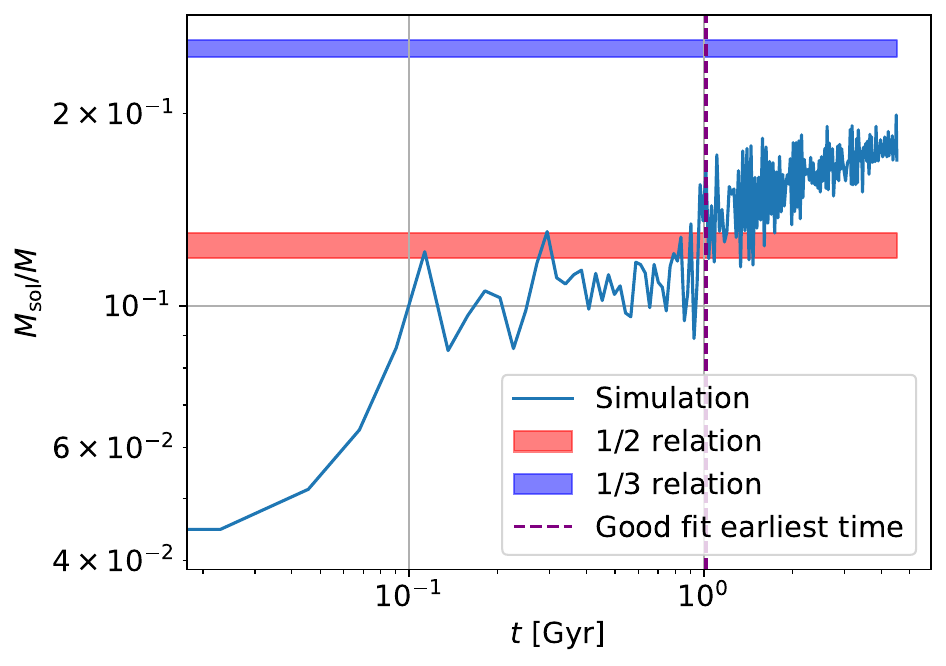}
\quad
\includegraphics[width=0.45\textwidth]{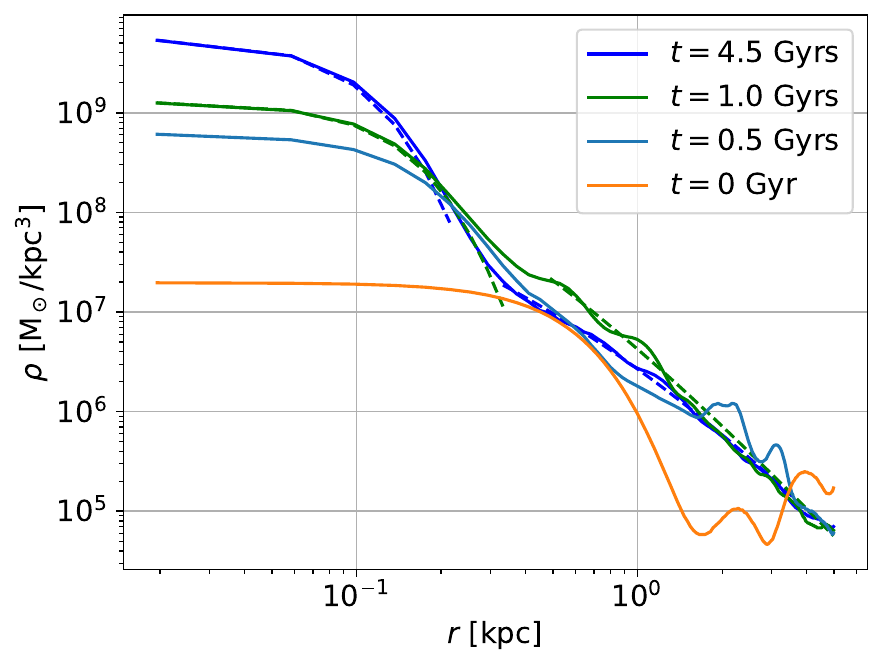}
    \caption{Results from a simulation with the {\it soliton merger (I)} initial conditions (in which the initial solitons have the same radius). The top four panels show the projected density field at $t=0,\,0.5,\,1.0,\,4.5$~Gyr, in $\unit{M_\odot\per\kilo\parsec\squared}$ units (logarithmic scale). The bottom-left panel shows the time-evolution of $M_{\rm sol}/M$. We find that the virialized soliton-halo system forms at $t_{\rm form}\approx \SI{1}{\giga\year}$ (dashed vertical line). Red and blue bands indicate the $1/2$ and $1/3$ relations, respectively. The width of these bands reflects the short-time fluctuations in $E_{\rm kin}$ around the final time (corresponding to the horizontal error bars in Fig.~\ref{fig:E_M}), which affects the value of $E_{\rm kin}/M^3 $ and hence the prediction of $M_{\rm sol}/M$. The bottom-right panel shows the radially averaged density profile around the center of the soliton at different times. The dashed lines show the soliton + NFW fit of the density profile, for the times $t=1.0,\,4.5$~Gyr at which this is good.
   }
    \label{fig:sol_halo_example}
\end{figure}

Fig.~\ref{fig:sol_halo_example} shows results from a sample simulation with \emph{Soliton merging (I)} initial conditions, which illustrates the formation of the soliton-halo system and motivates our procedure for when this is first identified. 
The top four panels show column density snapshots at $t=0,\,0.5,\,1.0,\,4.5$~Gyr. The bottom-left panel shows the time evolution of $M_{\rm sol}/M$, with $M_{\rm sol}$ estimated from the peak density $\rho_0$ using the soliton relation $ M^4_{\rm sol} = 64\pi\rho_0 / (G^3 m^6)$. The bottom-right panel shows the radially-averaged density profile at the corresponding times, together with the best-fit curves for the soliton and the NFW halo.  Based on the criterion above, the soliton-halo system forms at $t_{\rm form}\approx \SI{1}{\giga\year}$ (dashed  
line). Note that if we omit the halo region ($r>4 r_{\rm c}$) from the soliton-halo condition, we find that a soliton-like overdensity appears already at an earlier time with $t_{\rm form}\approx0.5$~Gyr. However, the top-right panel of Fig.~\ref{fig:sol_halo_example} shows that the halo is still evolving fast at that time. Our choice of procedure, including the choice of numerical factors, detailed in Appendix~\ref{s:sol_form_criterium}, is guided by qualitative tests of this kind.  The mass of the soliton 
is then recovered via the fitted density profile.\footnote{Refs.~\cite{Chan:2022bkz} pointed out that fitting the density profile with the soliton form can cause an upward bias in the inferred mass of the soliton component of the field, due to interference with unbound traveling wave fluctuations. Simple estimates suggest that this effect can range between ten percent to about a factor of two. We do not aim at a  better accuracy, nor believe this is required for current phenomenological applications.}

Solitons exhibit quasi-normal mode fluctuations~\cite{Guzman2004}, along with secular mass accretion from the surrounding halo~\cite{Levkov:2018kau,Eggemeier:2019jsu,Chen:2020cef,Chan:2022bkz}; 
both of which are apparent in the bottom-left panel of Fig.~\ref{fig:sol_halo_example}. Because of these effects, any single numerical value for $M_{\rm sol}$ would only capture part of the physics.  
Instead, when analyzing results, we report a range of $M_{\rm sol}/M$ values for each simulation, from $t = t_{\rm form}$ to the final simulation time $t_{\rm final}$, see e.g. Fig.~\ref{fig:E_M}. 
$E_{\rm kin}$ also undergoes fluctuations, which we represent as horizontal error bars on $E_{\rm kin}/M$, which we extract from the range of values around the final time. 

We end simulations when the soliton growth is deep in the era of secular growth\footnote{In this regime, the soliton mass accretion follows a power law behavior with time, $M_{\rm sol} \sim t^n$ with $n <1$. This behavior can be seen from the bottom-left panels of Fig.~\ref{fig:sol_halo_example} and Fig.~\ref{fig:sample_run}. A detailed study of the soliton mass accretion is beyond the scope of this work. Studies of the soliton mass growth include Refs.~\cite{Levkov:2018kau,Eggemeier:2019jsu,Chen:2020cef,Chan:2022bkz}.}, such that the largest value of $M_{\rm sol}$ reached is not very sensitive to the (somewhat arbitrary) value of $t_{\rm final}$. This is due to the slow mass accretion rate during this era, and also because the growth of the soliton in finite-box simulations is bounded above, in a sense that we explain further below. The specific value of $t_{\rm final}$ chosen depends on the particular initial conditions. For example, for the background initial conditions we ensure that $t_{\rm final} \gg t_{\rm rel}$, where $t_{\rm rel}$ is the relaxation time (see Ref.~\cite{Levkov:2018kau}); for soliton merging runs, collapse etc. we ensure that $t_{\rm final} $ is much greater than the free-fall time $t_{\rm ff}$, estimated as $t_{\rm ff} \sim 0.5 (GM/L^3)^{-1/2}$.

\section{Core-halo relation: Results} \label{s:results}

In Fig.~\ref{fig:E_M}, we summarize the soliton-halo relation obtained from all simulations and initial conditions. We express this in terms of the energy-to-mass ratios, $E/M$, for both the soliton and the halo. For a realistic system, the halo may extend beyond the volume that we simulate; however, as discussed below, $E/M$ is relatively unaffected by the cutoff set by the box. 
To avoid potentially unphysical results  from  the undetermined constant in the potential energy, we use the kinetic energy  $E_{\rm kin}$, 
instead of the total energy $E_{\rm tot}$.\footnote{See discussion in Section~IIA of~\cite{Bar:2019bqz}. Given this undetermined constant, plotting results in terms of $E_{\rm tot,(sol)}$ may introduce confusion when comparing different groups.}
This is justified under the assumption that the system is virialized. The soliton mass $M_{\rm sol}$  and kinetic energy $E_{\rm kin,sol}$  are obtained from fitting its density profile.  
Finally, the red line in Fig.~\ref{fig:E_M} shows the kinetic energy version of the $1/2$ relation, Eq.~(\ref{EMEM}).  
Over the time-range between $t_{\rm form}$ and $t_{\rm final}$, the data points fall around the $1/2$ relation.

To compare with existing literature and other soliton-halo relations, in particular the $1/3$ relation, Eq.~(\ref{eq:13}), we follow Refs.~\cite{Mocz:2017wlg,Zagorac:2022xic} in defining the dimensionless variable 
\begin{equation}\label{eq:Xi}
\Xi \equiv \frac{|E_{\rm tot}|}{M^3} \frac{1}{G^2 m^2} \ .
\end{equation}
This is an invariant of the Schr\"odinger--Poisson system under the residual rescaling symmetry described in Appendix~\ref{s:code} and is thus a natural candidate to enter in a soliton-halo relation. 
A class of soliton-halo relations, including the $1/2$ and $1/3$ relations, can be parametrized as
\begin{equation} \label{eq:sol_halo_rel}
\frac{M_{\rm sol}}{M} = \alpha\,\Xi^\beta \  .
\end{equation}  
In the literature, $\alpha$ and $\beta$ have been fitted as independent parameters. However, requiring the relation to hold in the limit $M_{\rm sol} \to M$ fixes $\alpha \simeq (0.054)^{-\beta}$ (see Appendix~\ref{s:soliton}). Perhaps not surprisingly, given that the fits to simulation data were obtained from systems with $M_{\rm sol}$ not much smaller than $M$, previously reported numerical values of $\alpha$ and $\beta$ are consistent with this limit. For example, the $\beta=1/2$ relation in Ref.~\cite{Schive:2014hza} gives $\alpha \approx 4.2$, while Ref.~\cite{Mocz:2017wlg} finds $\beta = 1/3$ and $\alpha \approx 2.6$.

\begin{figure}
	\centering
\includegraphics[width=0.7\textwidth]{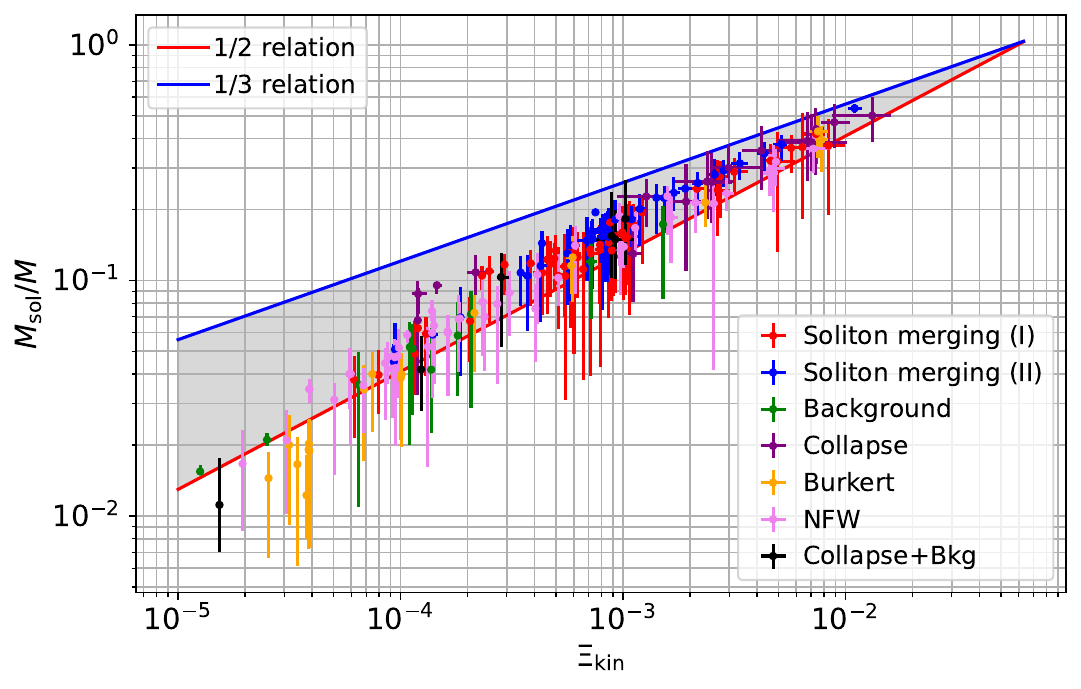}
	\caption{The ratio $M_{\rm sol}/M$ between the soliton and halo masses obtained in numerical simulations, as a function of the dimensionless Schr\"odinger--Poisson invariant $\Xi=|E|/(M^3G^2m^2)$, 
    where $E$ is calculated from the kinetic energy using the virial theorem. The red and blue lines show the $1/2$ and $1/3$ relations, respectively. Once the soliton-halo system is formed, it typically lies within (or at most slightly below) the band bracketed by these two relations, shaded gray. Error bars reflect variations in the soliton-halo relation over the system's evolution (as $M_{\rm sol}/M$ increases) and the impact of soliton fluctuations. 
    }
\label{fig:generic_final_snapshot}
\end{figure} 

\begin{figure}
    \centering    \includegraphics[width=0.7\linewidth]{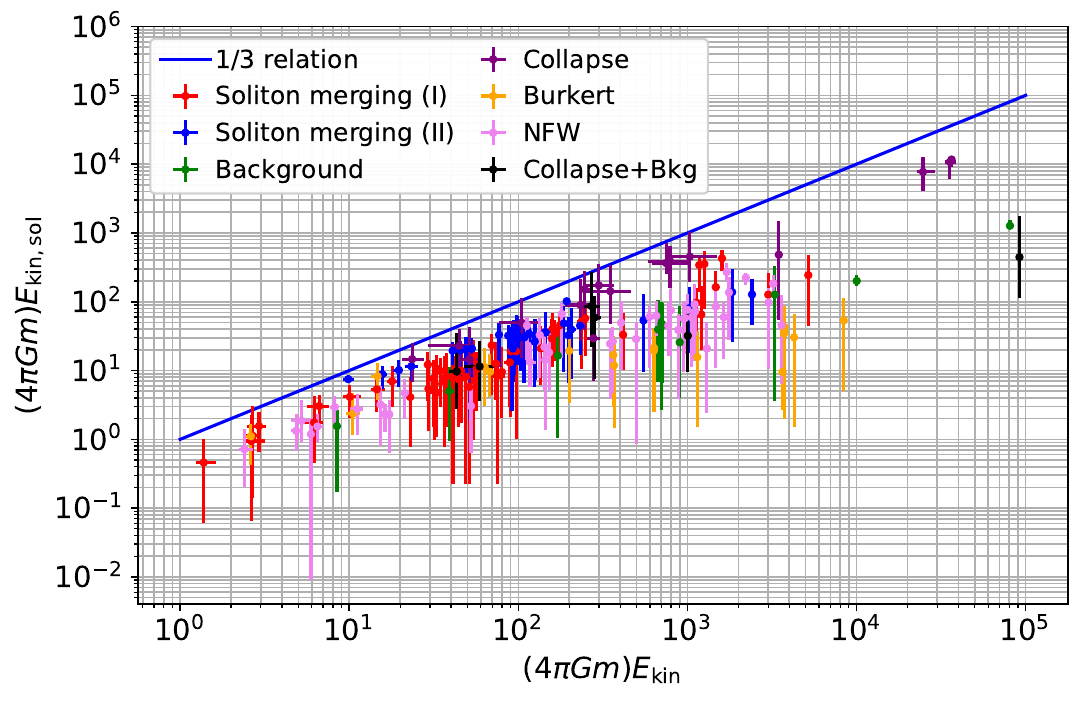}
    \caption{The soliton kinetic energy, $E_{\rm kin,sol}$, compared to the total kinetic energy in the simulation box, $E_{\rm kin}$ (which approximates the halo's plus soliton's kinetic energy). We also show the line $E_{\rm kin,sol} = E_{\rm kin}$. As expected, $E_{\rm kin,sol} < E_{\rm kin}$ in all cases, which is consistent with all data points falling below the $1/3$ soliton-halo relation in Fig.~\ref{fig:generic_final_snapshot}.}
    \label{fig:ekin_ekin}
\end{figure}

In Fig.~\ref{fig:generic_final_snapshot}, we present the same results as in Fig.~\ref{fig:E_M}, but using the variables $M_{\rm sol}/M$ and $\Xi$.  
As before, we replace $|E_{\rm tot}|$ in Eq.~(\ref{eq:Xi}) by  $E_{\rm kin}$. 
Consistently with Fig.~\ref{fig:E_M}, our results for $M_{\rm sol}/M$ fall around the $1/2$ relation. 
They also lie below the $1/3$ relation, with the trend becoming clearer at small $M_{\rm sol}/M$. This agrees with our expectation, discussed in Section~\ref{s:intro}, that the $1/3$ relation is an upper bound. To illustrate this point further, in Fig.~\ref{fig:ekin_ekin} we plot $E_{\rm kin,sol}$ versus the total kinetic energy $E_{\rm kin}$, i.e. the kinetic version of Eq.~(\ref{eq:13_simple}), confirming $E_{\rm kin,sol} < E_{\rm kin}$. 
Note that the separation between the values of $M_{\rm sol}/M$ corresponding to the $1/2$ and $1/3$ relations is larger for smaller $\Xi$. Consequently, if a soliton forms near the $1/2$ line, it takes longer to evolve toward the $1/3$ line. This may account for the clustering of data points near $1/2$ at lower values of $\Xi$ in Fig.~\ref{fig:generic_final_snapshot}.

\begin{figure}
    \centering
\includegraphics[width=0.54\linewidth]{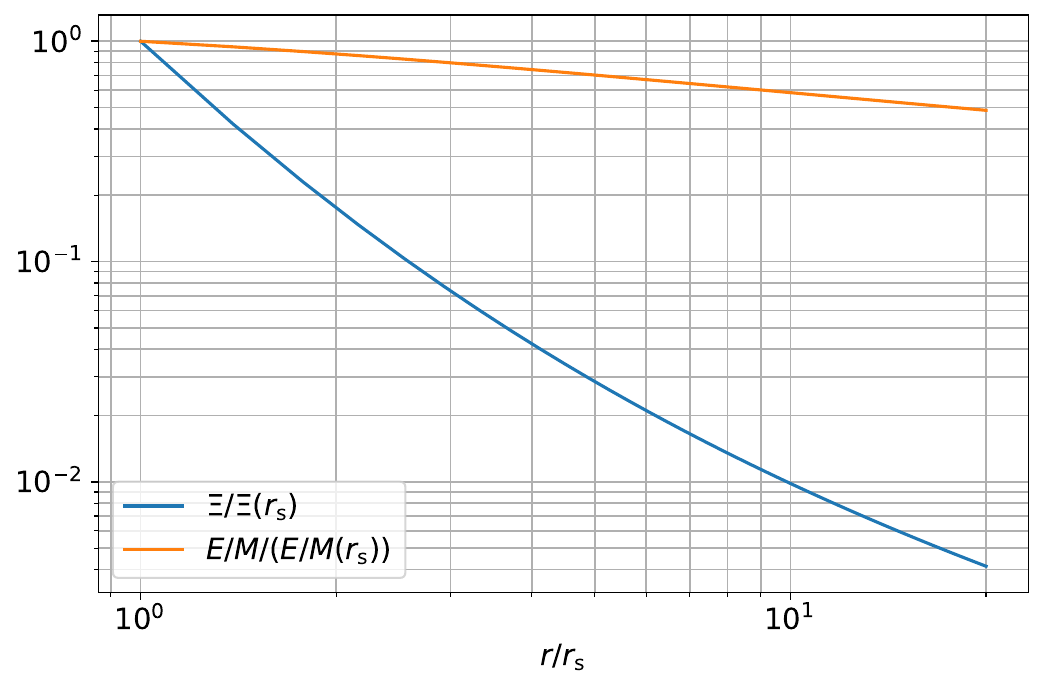}
    \caption{The radial dependence of the Schr\"odinger--Poisson invariant $\Xi(r)\propto|E(r)|/M(r)^3
    $ and of $|E(r)|/M(r)$ in an NFW halo with scale radius $r_{\rm s}$, normalized to their values at $r=r_{\rm s}$. These are calculated using the analytic form of the NFW density profile and gravitational potential. $\Xi(r)$ has a much stronger dependence on $r$ than $|E(r)|/M(r)$, and therefore when evaluated in simulations is more sensitive to the finite-box size in which a soliton-halo system is contained. 
    }
    \label{fig:Xi_EM_NFW}
\end{figure}

Despite its convenience in formulating the soliton-halo relation and describing simulation results, $\Xi$ is unfortunately badly affected by finite-volume effects.  
On the other hand, the quantity $E/M$ used in the relation in Eq.~(\ref{EMEM}) is much less sensitive to such artifacts. To illustrate this, in Fig.~\ref{fig:Xi_EM_NFW}  we show the $r$-dependence of $|E(r)|/M(r)$ and $\Xi(r)\propto|E(r)|/M^3(r)$ for an NFW halo of scale radius $r_{\rm s}$ by integrating $E$ and $M$ separately from the halo center to the point $r$. This procedure mimics how such variables vary as a function of the box size in finite-volume simulations that do not extend to the virial radius of a simulated halo system. $E(r)/M(r)$ only changes by a factor of 2 as $r$ varies from $r_{\rm s}$ to $10 r_{\rm s}$, however $\Xi(r)$ varies by more than two orders of magnitude. 

\begin{figure}
    \centering

\includegraphics[width=0.31\linewidth]{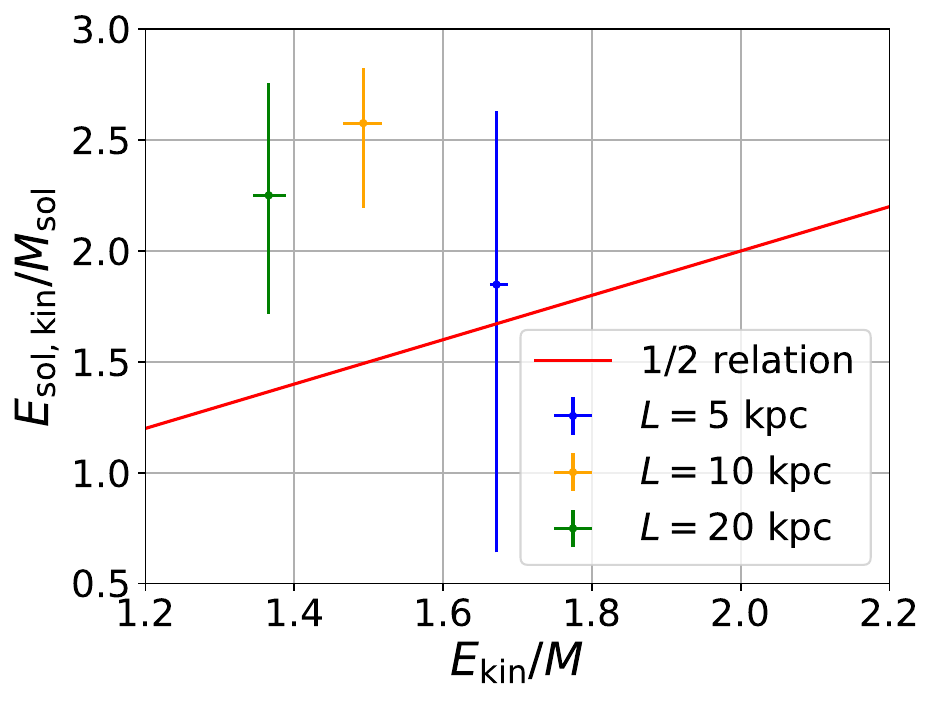} \ \ 
\includegraphics[width=0.33\linewidth]{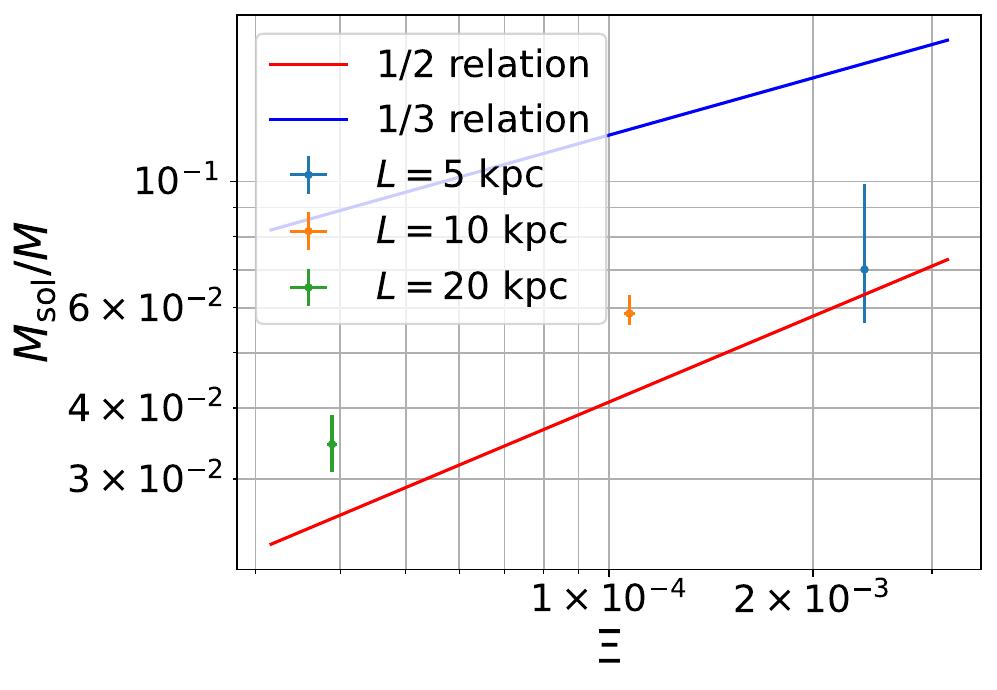}\ \ \ \ \ \ 
\includegraphics[width=0.295\linewidth]{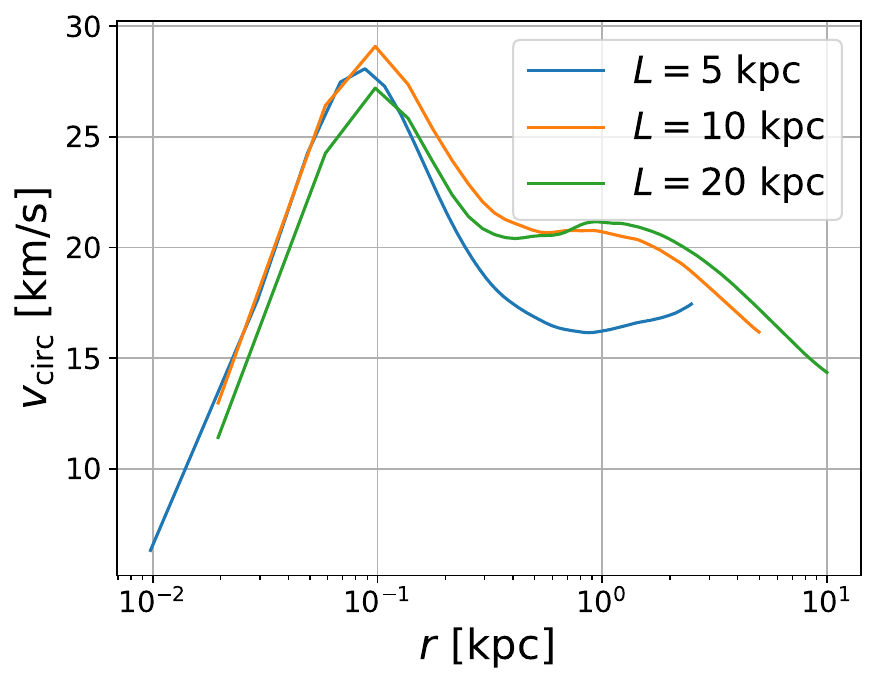} 
    \caption{The soliton-halo relation expressed in terms of $E/M$ (left panel) and $\Xi$ (center panel), arising from simulations with the same halo initial conditions but in different sized boxes, with box length $L$. The soliton-halo relation shows significantly less sensitivity to $L$ when expressed in terms of $E/M$ rather than $\Xi$. We also show the circular velocity $v_{\rm circ}(r)$ in the right panel. }
    \label{fig:vel_diff_L}
\end{figure}

As further illustration of the weak dependence of $E/M$ on the box length $L$, in Fig.~\ref{fig:vel_diff_L} we show results from three simulations for $m=10^{-21}\,\unit{\electronvolt}$ starting from NFW halo initial conditions with same central density and scale radius $r_{\rm s} = \SI{0.5}{\kilo\parsec}$, but different $L=5,\,10,\,20$~kpc. A prominent soliton forms, with its mass varying by at most 10\% across the three box sizes. 
The left and center panels compare the soliton-halo relations using $E/M$ and  $\Xi$, respectively. $E_{\rm kin}/M$ and $E_{\rm sol,kin}/M_{\rm sol}$ remain clustered within 20\% across the three cases, whereas $\Xi$ and $M_{\rm sol}/M$ vary by a factor of 2 and 5, and thus are much more sensitive to $L$.

Importantly for observational comparisons, a soliton-halo system that lies around the $1/2$ relation leads to a velocity rotation curve $v_{\rm circ}(r)\equiv\sqrt{GM(r)/r}$ that has two distinct and approximately equal peaks, at distances of the order of the soliton radius and the halo scale radius, respectively~\cite{Bar:2018acw}.  
The soliton peak predicted from simulations is largely unaffected by finite-volume effects. Indeed, $v_{\rm circ}(r)$ only depends on the mass distribution at a distance smaller than $r$ (for an approximately spherical system), and is thus sensitive to finite box size effects only to the extent that these alter the density distribution of the system within the simulated volume. To demonstrate this, in the right panel of Fig.~\ref{fig:vel_diff_L} we plot the rotation curves in the three simulations in different size boxes described above. The amplitude and position of the inner peak, associated to the soliton, is the same within $20\%$ in all cases.  
The outer peak, linked to the halo, is instead sensitive to box size if this is too small; for instance, it is absent in a $\SI{5}{\kilo\parsec}$ box.

\section{Discussion} \label{s:discussion}

\subsection{Consistency with the evaporation lower limit of Chan et al. 2022
} \label{s:theory}

Chan, Sibiryakov, and Xue~\cite{Chan:2022bkz}  conducted a perturbative analysis of soliton mass absorption and evaporation in a statistically-homogeneous background with a Maxwellian velocity distribution and characteristic momentum  
$k_{\rm bkg}$, as given by Eq.~(\ref{eq:cov}). They demonstrated that a soliton undergoes evaporation due to gravitational scattering when the background particles are significantly faster than those in the soliton. More precisely, this occurs if~\footnote{This is obtained combining Eqs.
~(2.18) and (2.12) of Ref.~\cite{Chan:2022bkz} with the condition $T_{\rm s}/T_{\rm g}\lesssim0.08$ obtained in the same reference.}
\begin{equation}\label{eq:crit_evap}
\frac{E_{\rm kin,sol}}{M_{\rm sol}} \lesssim 0.12 \,
\frac{k^2_{\rm bkg}}{2m^2} \ .
\end{equation} 
In the opposite limit, the soliton instead grows by accreting mass from the background. 

The $1/2$ relation of Eq.~(\ref{EMEM}), phrased in terms of kinetic rather than total energy, can be written in a form similar to Eq.~\eqref{eq:crit_evap}: 
\begin{equation}\label{eq:crit_evap1}
\frac{E_{\rm kin,sol}}{M_{\rm sol}} = \frac{E_{\rm kin}}{M} =\frac{3k^2_{\rm bkg}}{ 4m^2}\, ,
\end{equation} 
where we used $E_{\rm kin}/M=3k_{\rm bkg}^2/ 4m^2$, valid for a Maxwellian distribution with variance $k_{\rm bkg}^2/2$, see Appendix~\ref{s:initial_cond}.

Comparing Eq.~(\ref{eq:crit_evap1}) with Eq.~(\ref{eq:crit_evap}), we obtain that solitons that just exceed the evaporation limit of Eq.~(\ref{eq:crit_evap}) are a factor 
$(3/(2\cdot0.12))^{1/2}\sim3.5$ lighter than prescribed by the $1/2$ relation, i.e. solitons evaporate if their mass is smaller than
\begin{equation}\label{eq:Msolevap}
    M_{\rm sol,\,evap} \simeq 1.2 \qty(\frac{|E_{\rm tot}|}{M} )^{1/2} \frac{1}{Gm} \ . 
\end{equation}

Although strictly valid only for a soliton embedded in a Maxwellian gas, Eqs.~\eqref{eq:crit_evap1} and~\eqref{eq:Msolevap} may still be applicable to a generic soliton-halo system, where $E_{\text{kin}}$, $M$, and $k_{\text{bkg}}$ are interpreted as the kinetic energy, mass, and characteristic momentum in an approximately virialized region that is larger than the soliton itself and containing a mass much larger than the soliton mass. 
In this case, only solitons with masses at most a factor of $\sim3.5$ below the $1/2$ line (red) in Fig.~\ref{fig:generic_final_snapshot} would be possible. The inhomogeneous nature of a halo and the fact that the particles in the halo are self-bound could affect this analysis. The evaporation limit of Eq.~(\ref{eq:Msolevap}) appears to nevertheless be approximately valid, see Fig.~\ref{fig:generic_final_snapshot}. 

\begin{figure}
    \centering 
\qquad    \includegraphics[width=0.4\textwidth]{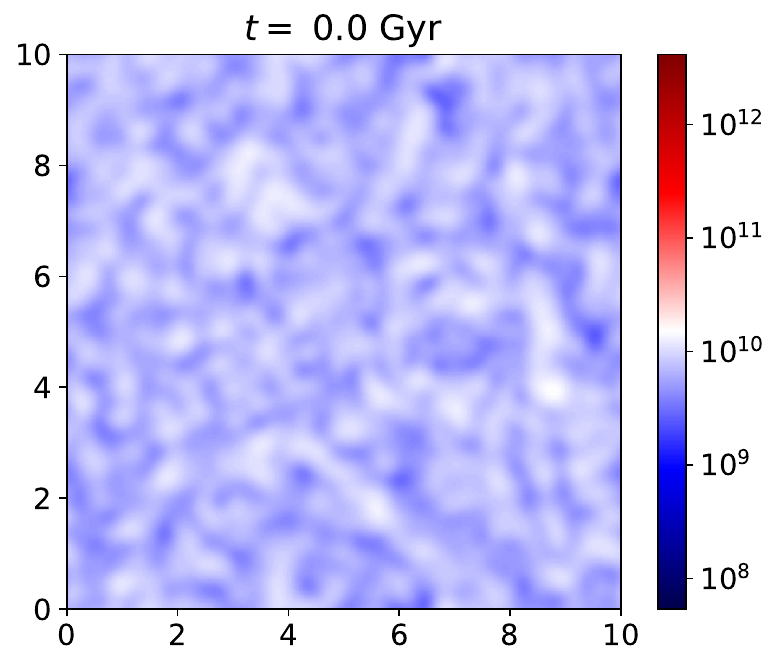}
\quad \includegraphics[width=0.4\textwidth]{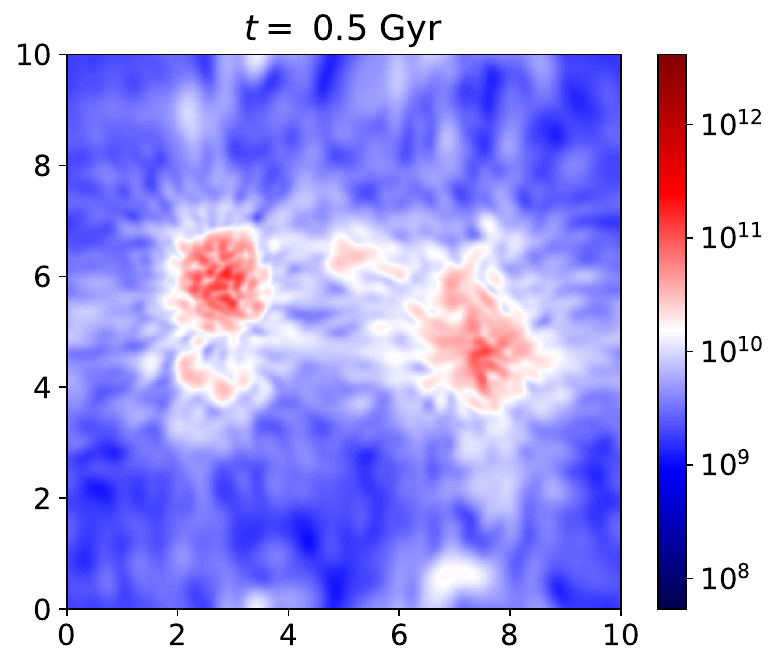}\\
\qquad \includegraphics[width=0.4\textwidth]{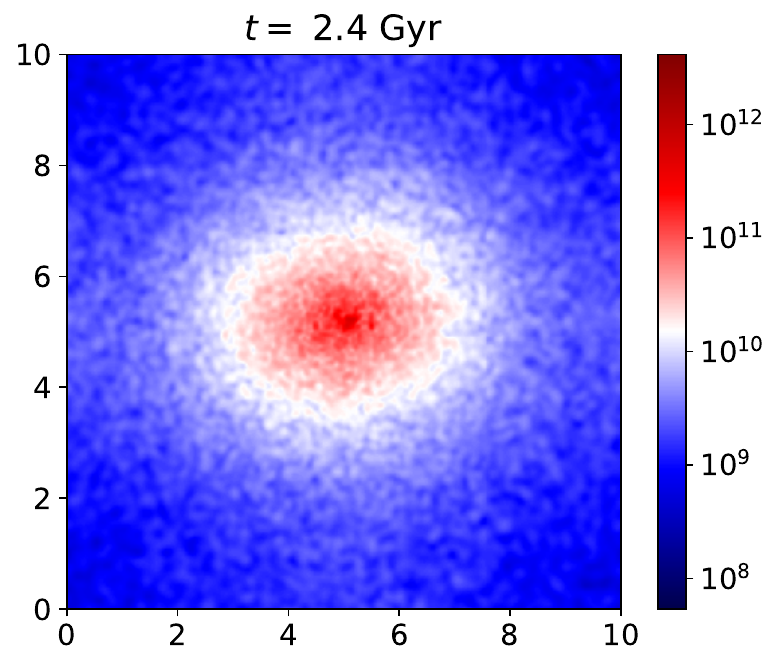}
\quad \includegraphics[width=0.4\textwidth]{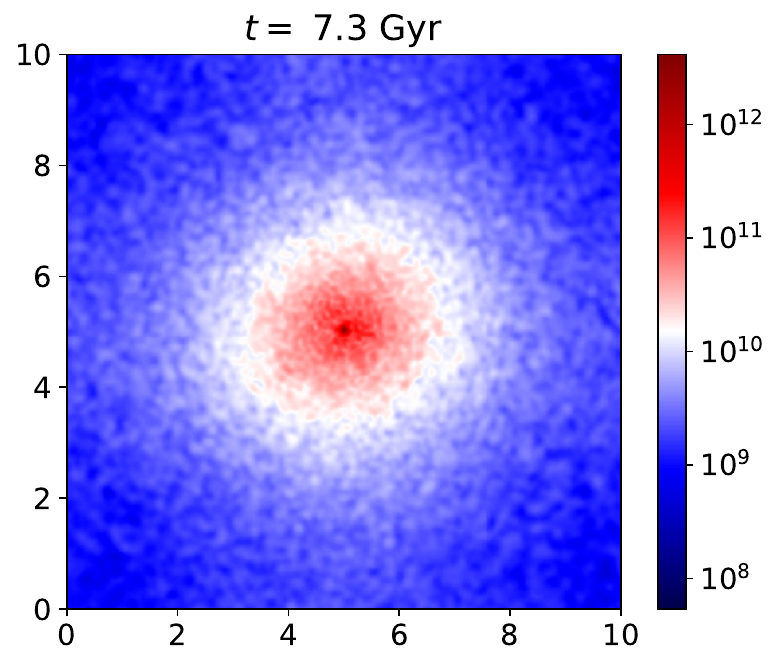}\\
\includegraphics[width=0.45\textwidth]{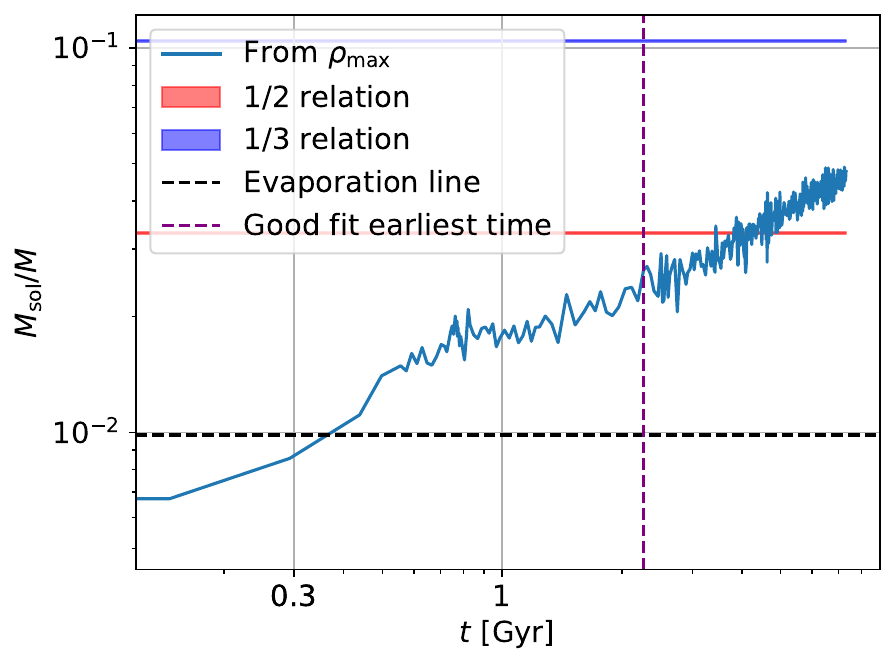} 
\quad \includegraphics[width=0.45\textwidth]{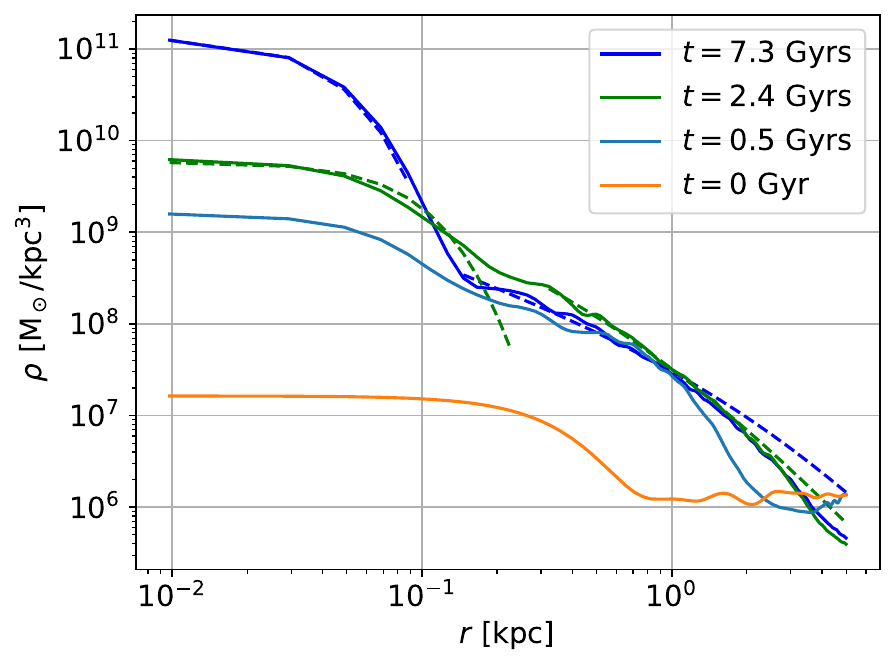}
    \caption{
    Formation the soliton-halo system and growth of the soliton in a typical simulation starting from initial conditions consisting of background noise, with Jeans wavelength $\lambda_{\rm J}/L \simeq 0.2$. The plots are as described in Fig.~\ref{fig:sol_halo_example}; in the bottom-left panel we additionally show the evaporation limit of Ref.~\cite{Chan:2022bkz}, below which a soliton is unstable to evaporation by scattering with the background. This limit is roughly a factor of $3.5$ smaller than the $1/2$ relation (in this plot the finite thickness of the $1/2$ and $1/3$ bands is hardly noticeable).
    }
    \label{fig:sample_run}
\end{figure}

Our results from simulations are consistent with such a limit.  An example of this behavior is plotted in Fig.~\ref{fig:sample_run}, which shows the growth of a soliton starting from background noise initial conditions (with {\it initial} $k_{\rm bkg}L =40$, $ \lambda_{\rm J}/L \simeq 0.2 $). 
Such initial conditions are fairly close to the simplifying assumptions in the analysis of Ref.~\cite{Chan:2022bkz}, except for the fact that $\lambda_{\rm J}/L <1$, which induces background collapse via Jeans instability and the formation of a self-bound halo. 
Note however that the evaporation limit should be calculated for the virialized halo, i.e. after the collapse of the initial background waves. 
During collapse $E_{\rm kin}$  becomes significantly larger than its initial value  $3k_{\rm bkg}^2 M / 4m^2$, leading to a tighter evaporation limit than that one would infer from the initial kinetic energy density. This is, approximately, taken into account by using the evaporation limit in the form of Eq.~(\ref{eq:Msolevap}). 
As shown in the bottom-left panel of Fig.~\ref{fig:sample_run}, when it forms, the soliton lies between the $1/2$ relation in Eq.~\eqref{eq:schive} (red) and the resulting evaporation limit in Eq.~\eqref{eq:Msolevap} (dashed black), both  computed using  $E_{\rm kin}$ from the simulation after soliton formation.

\subsection{The  $1/3$  relation is an upper bound
} \label{s:theory2}

Since the $1/3$ relation simply amounts to Eq.~(\ref{eq:13_simple}), it can at most set an upper bound on the soliton mass in a cosmological context, or in flat-space simulations where all field components are bound (i.e., have negative energy). In a finite volume, there could be unbound positive-energy debris that orbit the box, which would allow the $1/3$ relation to be violated if $E_{\mathrm{tot}}$ and $M$ are interpreted as the halo’s energy and mass (in an infinite volume, such debris would escape to infinity). However, by analyzing the spectrum of the ULDM field, see Appendix~\ref{s:spectrum}, we find that our simulations have a negligible quantity of positive energy debris. 
Figs.~\ref{fig:generic_final_snapshot} and~\ref{fig:ekin_ekin} are consistent with this point and thus the $1/3$ relation. Our presentation of kinetic rather than total energy in these figures avoids ambiguities with respect to the definition of total energy that would affect the soliton-halo relation; these however need to be resolved for direct comparison to other works in the literature, notably Ref.~\cite{Mocz:2017wlg}, which makes use of the total energy. We return to this point in Section~\ref{s:comparison}.

\subsection{Observational implications} \label{s:halos}

Given sufficient time, halos in our simulations grow a central soliton with an initial mass lying near the $1/2$ relation that then slowly increases in mass towards the $1/3$ relation. As we discuss further in Section~\ref{s:comparison}, we expect similar scaling for cosmological halos. 
A soliton-halo relation allows predictions to be made for observable rotation and dispersion velocity curves~\cite{Bar:2018acw}. 
A soliton of a given mass $M_{\rm sol}$ induces a characteristic peak in the circular velocity profile at radius $r_{\rm peak}$ with value $v_{\rm peak}$, which are given by
\begin{equation} \label{eq:vpeak}
v_{\rm peak} \approx 8.3 \qty(\frac{m}{10^{-21} \unit{\electronvolt}}) \qty(\frac{M_{\rm sol} }{ 10^7 M_{\odot}}) \unit{\kilo\meter\per\second} \ , \ \ \ r_{\rm peak} \approx 0.46 \qty(\frac{m}{10^{-21} \unit{\electronvolt}})^{-2} \qty(\frac{M_{\rm sol} }{ 10^7 M_{\odot}})^{-1} \unit{\kilo\parsec} \ .
\end{equation}
For a density profile consisting of a soliton embedded in a host halo and not too far from the $1/2$ relation, a doubly-peaked rotation curve is predicted. The inner peak is approximately given by Eq.~(\ref{eq:vpeak}), and is at radius  $r\sim2r_{\rm c}$~\cite{Bar:2018acw}, and the outer peak is induced by the NFW halo at radius $r\sim2r_{\rm s}$, with $r_{\rm s}$ the NFW scale radius. As the mass of a soliton increases above the $1/2$ relation, the relative amplitude of the second peak decreases.  
If the radial scale $r\sim 2r_{\rm c}$ can be resolved by observations (recall that $r_{\rm c} \propto 1/m$), the presence or lack thereof of the inner peak can constrain, or give evidence for, ULDM with a particular $m$.

\begin{figure}
    \centering
\includegraphics[width=0.5\linewidth]{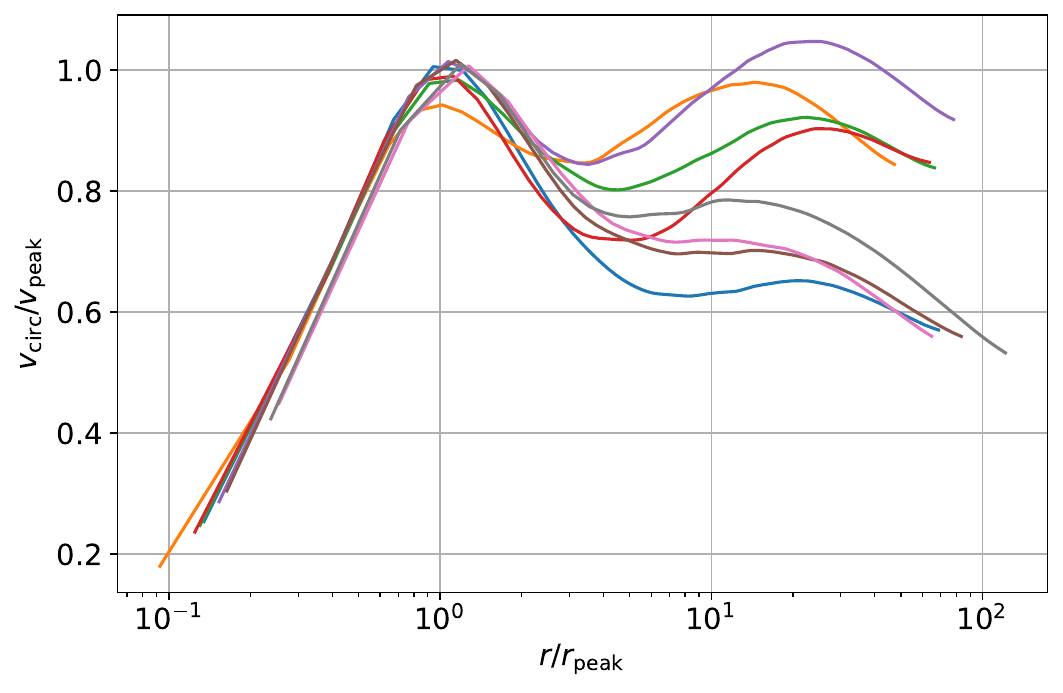}
    \caption{Velocity rotation curves for simulations with halo initial condition (Burkert and NFW), evaluated at $t\approx\SI{10}{\giga\year}$. Here, $r_{\text{peak}}$  is the distance at which the circular velocity $v_{\text{circ}}$  reaches its maximum for a soliton density profile matched to the central density in the simulations, and  $v_{\text{peak}}$ is the corresponding velocity. 
    The rotation curves plotted are obtained from the radially averaged density profile in the simulations (and hence $v_{\rm circ}$ is not exactly equal to $v_{\rm peak}$ at $r=r_{\rm peak}$).
    }
    \label{fig:rot_vel_10gyr}
\end{figure}

Fig.~\ref{fig:rot_vel_10gyr} shows the circular velocity curves of soliton-halo systems in simulations with halo initial conditions (NFW and Burkert). 
These are calculated from the radially-averaged density profiles, and normalized to $v_{\rm peak}$ and $r_{\rm peak}$, which are defined in Eq.~(\ref{eq:vpeak}) and evaluated using the soliton density profile fit.  
These results are plotted at a physical time of $10$~Gyr, and the masses and size of the halos are taken in the range relevant to dwarf galaxies and ULDM particle mass $m$ between $5\times10^{-22}\, \unit{\electronvolt}$ and $10^{-21} \,\unit{\electronvolt}$. We vary the box length between $10\, \unit{\kilo\parsec}$ and $40\, \unit{\kilo\parsec}$, and the halo parameters within $\rho_0 = 10^6 - 10^8 \,\unit{M_\odot\per\kilo\parsec\cubed}$, $r_{\rm s} = 0.5-1.5 \,\unit{\kilo\parsec}$. With such parameters, the ratio $r_{\rm s}/r_{\rm c}$ varies over $10 \lesssim r_{\rm s}/r_{\rm c} \lesssim 30$, as can be inferred from the position of the second peak in Fig.~\ref{fig:rot_vel_10gyr}.

Doubly-peaked rotation curves are generally {\it not} observed in the kinematical data of DM-dominated disk galaxies. This led Refs.~\cite{Bar:2018acw,Bar:2019bqz,Bar:2021kti} to infer a bound $m\gtrsim10^{-21}$~eV by assuming the $1/2$ soliton-halo relation. 
As this relation is consistent with our results for $m \simeq 10^{-21}$ eV in dwarf galaxy-like systems with ages near the Hubble time, our simulations support such a lower bound on $m$.

\subsection{Comparison with previous works} \label{s:comparison}

\subsubsection{Comparison with flat space-time simulations}
We now compare with Ref.~\cite{Mocz:2017wlg} and examine the consistency of our results with their claimed $1/3$ relation. Our numerical methods (i.e. our algorithm to solve the Schr\"odinger--Poisson equations) and the setting (flat space-time with periodic boundary) are similar to this reference. Moreover, as far as we could tell, our {\it soliton merging (II)} initial conditions match their initial conditions. Yet our results, as shown in Fig.~\ref{fig:generic_final_snapshot}, lie systematically below the $1/3$ relation, in apparent contrast to those of~\cite{Mocz:2017wlg}.

One possible reason for this discrepancy is the choice of Ref.~\cite{Mocz:2017wlg} to use the variable  $\Xi$  calculated from the total energy $E_{\rm tot}$ rather than the kinetic energy. Since, as mentioned, in a finite volume the potential energy $E_{\rm pot}$ contains an arbitrary additive constant, the total energy does too. If not chosen such that the potential energy matches its infinite volume value, such a constant means that e.g. the system would usually not obey the virial theorem $E_{\rm kin}=-E_{\rm tot}$.\footnote{Fig.~1 in Ref.~\cite{Mocz:2017wlg} shows the total, potential, and kinetic energy in their simulation. That plot shows that the virial theorem is indeed violated at the $\mathcal{O}(1)$ level. Digitizing the plot we find $2E_{\rm kin}/|E_{\rm pot}|\approx1.4$, instead of 1. This may not come as a surprise, since indeed in Ref.~\cite{Mocz:2017wlg}, the arbitrary constant is determined to make the average of $\Phi$ in the box equal to zero, as also our periodic boundary conditions solver does, see Appendix~\ref{s:code}. We thank P. Mocz for clarification of this point.}

\begin{figure}
    \centering
\includegraphics[width=0.48\linewidth]{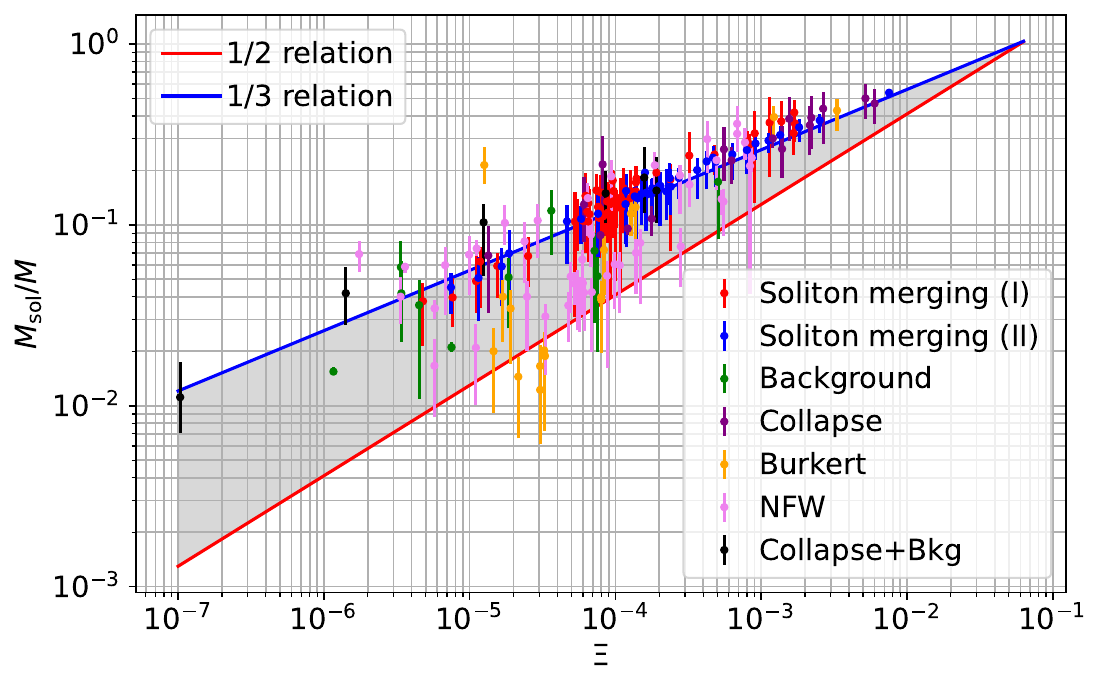}
\includegraphics[width=0.4\linewidth]{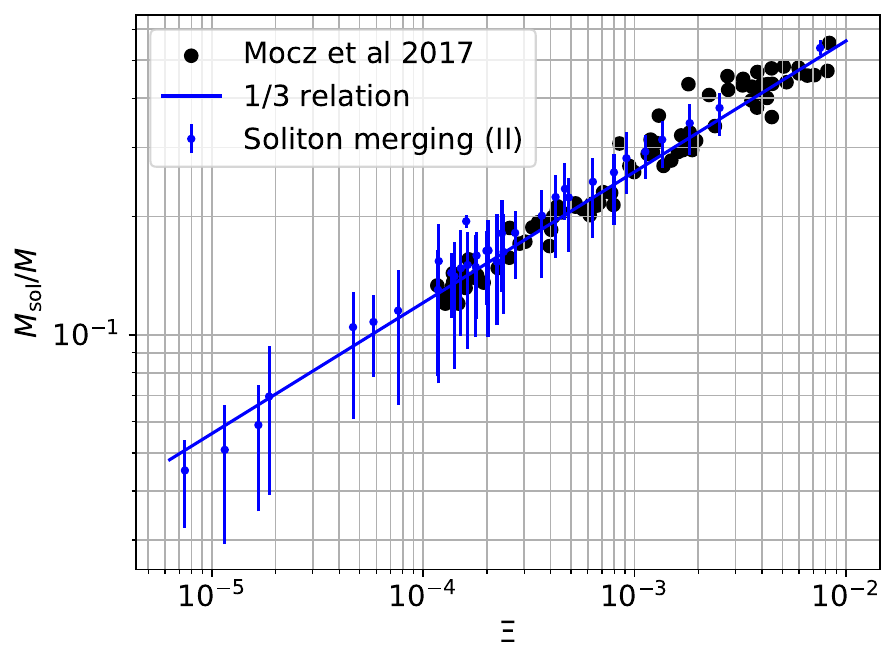}
    \caption{Left: the same as Fig.~\ref{fig:generic_final_snapshot}, but with $\Xi$ calculated using the total (instead of kinetic) energy. In the total energy, the potential energy is evaluated directly in finite volume, 
    i.e. without choosing the arbitrary constant to match the infinite-volume limit
    . Right: the same as the left plot, but for direct comparison with Ref.~\cite{Mocz:2017wlg}, we show only our results for {\it soliton merger (II)} initial conditions (blue markers), as used in Ref.~\cite{Mocz:2017wlg}. Digitized results from Ref.~\cite{Mocz:2017wlg} are shown by black dots.}
    \label{fig:etot_check}
\end{figure}

To check that this is the origin of our discrepancy with Ref.~\cite{Mocz:2017wlg}, in the left panel of Fig.~\ref{fig:etot_check} we plot our results in terms of $\Xi$ calculated using the total energy without any correction in $E_{\rm pot}$. 
With this prescription our results cluster around the $1/3$ relation, slightly exceeding it in some cases. 
Meanwhile, in the right panel we plot only data from {\it soliton merger (II)} initial conditions (blue markers), which are the direct parallel of the simulations of Ref.~\cite{Mocz:2017wlg}. The results obtained in this way are fully consistent with those of~\cite{Mocz:2017wlg}; for comparison, we show also digitized data from Fig.4 {\it there} (black points). 

\begin{figure}
    \centering
\includegraphics[width=0.45\linewidth]{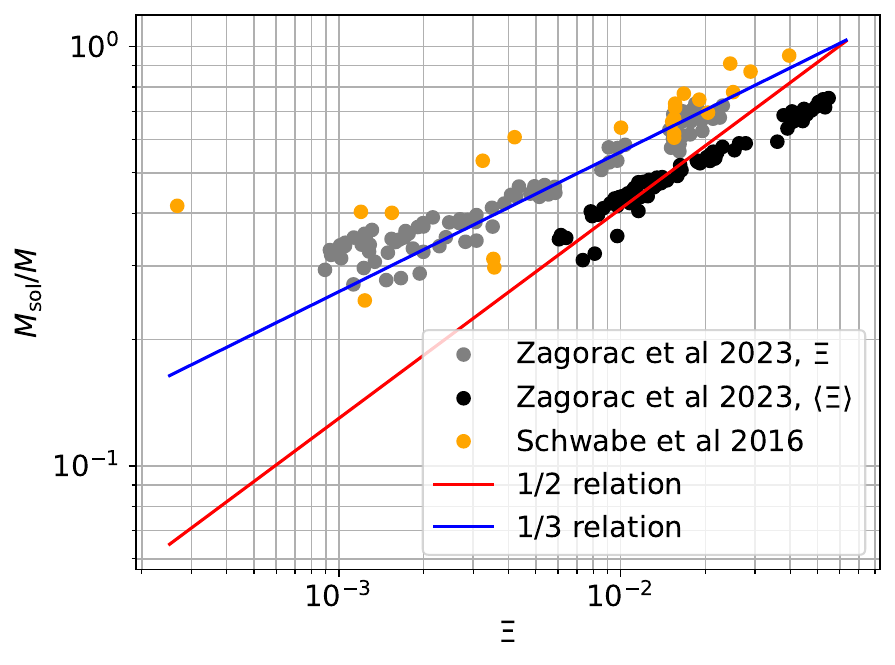}
    \caption{Simulation results for the soliton halo relation from Ref.~\cite{Zagorac:2022xic} and Ref.~\cite{Schwabe:2016rze}, compared with the $1/2$ and  $1/3$ relations. }
    \label{fig:zag}
\end{figure}

Other works in the literature used different simulation setups to analyze the soliton-halo relation, so only indirect comparisons 
are possible.  
Ref.~\cite{Zagorac:2022xic} used absorbing boundary conditions. Since, in this case, $\Xi$ is not constant in time, Ref.~\cite{Zagorac:2022xic} presented results both with respect to the initial value of $\Xi$, and with respect to a short time average $\langle\Xi\rangle$. In Fig.~\ref{fig:zag} we superimpose the $1/2$ and  $1/3$ relations alongside digitized results from that paper. 
We do the same for the results in Ref.~\cite{Schwabe:2016rze}, which also used absorbing boundary conditions, again in Fig.~\ref{fig:zag}.\footnote{We could not determine whether $\Xi$ used in that reference was calculated from the initial data, or another procedure as in~\cite{Zagorac:2022xic}.} Speculatively, if these references did use an unadjusted $E_{\rm pot}$, that might (possibly partially) account for those data points clustered around and above the $1/3$ line. A direct comparison by re-analyzing such data using $E_{\rm kin}$ may be useful.

\subsubsection{Comparison with cosmological simulations} \label{s:cosmo_comparison}
In a cosmological context, the soliton-halo relation is usually written as
\begin{equation} \label{eq:sol_halo_cosmo}
a^{1/2}M_{\rm sol} = A_{\rm h} \qty(\sqrt{\frac{\zeta(a)}{\zeta(1)}} M)^{B_{\rm h}}\ ,
\end{equation}
where $M_{\rm sol}$ and $M$ are the mass of a soliton and halo respectively, and $A_{\rm h}$ and $B_{\rm h}$ are parameters of the relation. Here, $a$ is the cosmological scale factor, and $\zeta(a)$ is the cosmology-dependent factor that relates the average overdensity of the halo within its virial radius, $r_{\rm vir}$,  
to the cosmological dark matter density 
at the time of formation~\cite{Schive:2014hza}, i.e.
\begin{equation}\label{eq:Mhalocosmo}
M \equiv \frac{4\pi}{3}\frac{\rho_{\rm m}}{a^3} r^3_{\rm vir} \,\zeta(a) \ ,
\end{equation} 
and $\rho_{\rm m}$ is the cosmological dark matter density today. Note that $\zeta(a) =(18\pi^2 + 82(\Omega_{\rm m}(a) - 1) - 39(\Omega_{\rm m}(a) - 1)^2)/\Omega_{\rm m}(a)$, with $\Omega_{\rm m}(a)$ the matter cosmological abundance at scale factor $a$~\cite{bryan1998statistical}.

We can attempt to match the general form of the soliton-halo relations we consider, Eq.~(\ref{eq:sol_halo_rel}),  with the cosmological form of Eq.~(\ref{eq:sol_halo_cosmo}) as follows. 
For a virialized halo, it is reasonable to assume
\begin{equation}
|E_{\rm tot}| \sim \frac{GM^2}{r_{\rm vir}}  \ .
\end{equation}
Plugging this into  Eq.~(\ref{eq:sol_halo_rel}) with $r_{\rm vir}$ expressed in terms of $M$ from Eq.~\eqref{eq:Mhalocosmo}
, leads to
\begin{equation}\label{eq:Msolcosmo}
a^\beta M_{\rm sol} = \alpha_\beta \left(\frac{4\pi \rho_{\rm m} \zeta(1)}{3G^3 m^6}  \right)^{\frac{\beta}{3}} \qty(\frac{\zeta(a)}{\zeta(1)})^{\beta-\frac12}  \qty(\sqrt{\frac{\zeta(a)}{\zeta(1)}}  M)^{1-\frac{4\beta}{3}} \implies \ \  B_{\rm h}(\beta) = 1-\frac{4\beta}{3} \ .
\end{equation}
We see that this only leads to a soliton-halo relation of the form of Eq.~(\ref{eq:sol_halo_cosmo}) for $\beta=1/2$, in which case $B_{\rm h}=1/3$ (this was also found in Ref.~\cite{Schive:2014hza}). On the other hand, for a generic $\beta$, the most we can do is neglect the effects of cosmology, by setting $a=1$, $\zeta =1$. With such an assumption,
we can relate the
parameters $A_{\rm h}$ for different $\beta$ as
\begin{equation} \label{eq:corr_cosmo}
A_{\rm h}(\beta) = \alpha_{\beta} \qty(\frac{A_{\rm h}(\beta')}{\alpha_{\beta'}})^{\beta/\beta'} \implies \ \  A_{\rm h}(1/3 ) \approx  A^{2/3}_{\rm h}(1/2) \ ,
\end{equation}
where we used $\alpha_{1/2}=4.2$ and $\alpha_{1/3}=2.6$.\footnote{To analyze such data without neglecting cosmological effects, we would require soliton and halo masses without cosmological normalizations, which have typically not been reported in the literature.}

\begin{figure}
    \centering
\includegraphics[width=0.6\linewidth]{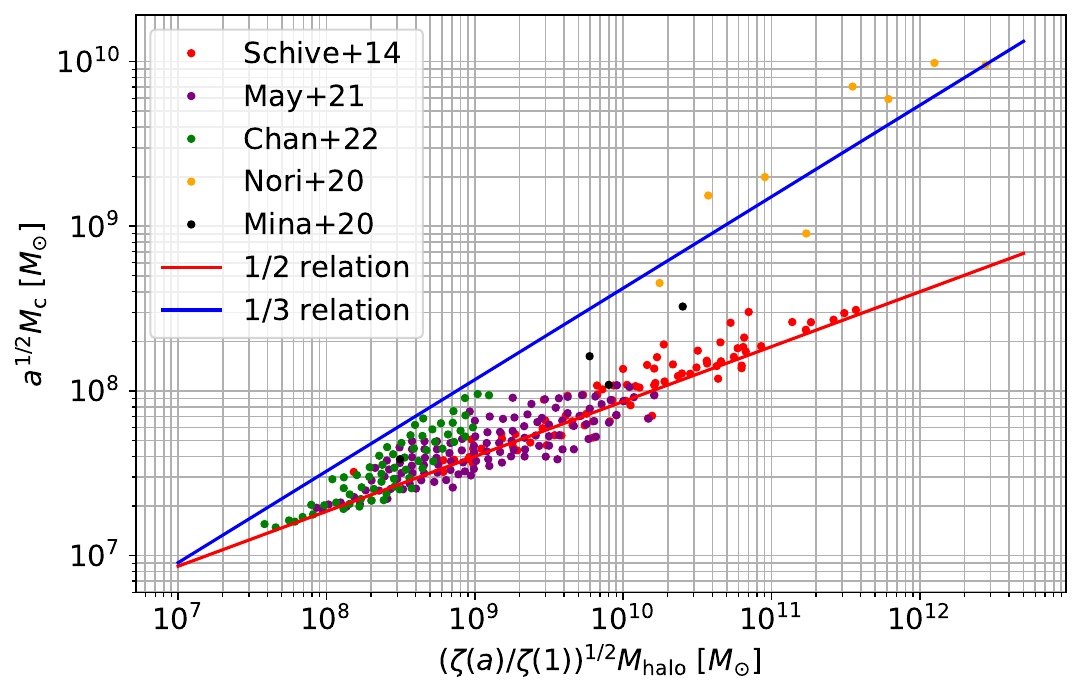} 
    \caption{Results from cosmological simulations~\cite{Schive:2014hza,May:2021wwp,Nori:2020jzx,Chan:2022bkz}, as collected by Ref.~\cite{Chan:2021bja}, plus from Ref.~\cite{Mina:2020eik}, scaled for an ULDM mass $m=\SI{8e-23}{\electronvolt}$ (results in Ref.~\cite{Mina:2020eik} are shown for $m=\SI{2.5e-22}{\electronvolt}$). Notice that here, results are expressed using $M_{\rm c} \simeq M_{\rm sol}/4.2$. (For Ref.~\cite{Mina:2020eik} we show only halos at $z=2.5$, and exclude their halo-4 that seems far from virialized, see Figs.~4 and~6 \emph{there}.) Also shown is the $1/2$ relation translated to a cosmological setting, and the  $1/3$ relation (the latter assumes no cosmology effects, see discussion in Sec.~\ref{s:cosmo_comparison}).
    }
    \label{fig:Chan_plot}
\end{figure}

Fig.~\ref{fig:Chan_plot} shows  results from cosmological simulations~\cite{Schive:2014hza,May:2021wwp,Nori:2020jzx,Chan:2022bkz} as collected by Ref.~\cite{Chan:2021bja}, together with some halos from Ref.~\cite{Mina:2020eik}. In Ref.~\cite{Mina:2020eik}, results from 5 halos at different redshift are shown. In Fig.~\ref{fig:Chan_plot}, we only show results from the halos at their latest redshift $z=2.5$ and exclude their halo-4; as one can evince from Fig.~4, Fig.~6 \emph{there}, such excluded halos may not be virialized. Ref.~\cite{Mina:2020eik} used $m=\SI{2.5e-22}{\electronvolt}$, whereas Ref.~\cite{Chan:2021bja} used $m=\SI{0.8e-22}{\electronvolt}$. According to the rescaling Eq.~\eqref{eq:rescaling}, Eq.~\eqref{eq:tildeM}, to translate the points of Ref.~\cite{Mina:2020eik} to $m=\SI{0.8e-22}{\electronvolt}$ while keeping the same time-scales, one should rescale $M \to (2.5/0.8)^{3/2} M $.  
We compare these with the $1/2$ cosmological relation of Eq.~\eqref{eq:Msolcosmo} with $\beta=1/2$, and also
with the $1/3$ relation in the form of Eq.~\eqref{eq:sol_halo_cosmo} neglecting cosmological effects (i.e. with $a=1$ and $B_{\rm h}(1/3)$ and $A_{\rm h}(1/3)$ as in Eqs.~(\ref{eq:Msolcosmo}) and~(\ref{eq:corr_cosmo})).  
We find rough agreement with the expectations that the $1/2$ relation is a rough lower bound and the $1/3$ relation is an upper bound, despite neglecting cosmological effects in the latter.\footnote{The results of Ref.~\cite{Nori:2020jzx} are outliers with respect to the other groups results we show. Although we do not have a clear explanation as of why this is the case, we note that Ref.~\cite{Nori:2020jzx} uses a smoothed particle hydrodynamic approach that emulate the ULDM wave effects, which may affect the soliton-halo relation obtained.}

\section{Summary}\label{s:sum}

The presence of cored, coherent over-densities (`solitons') inside halos is a key prediction of ULDM theories. It is important to understand how massive such solitons are in order to compare predictions of ULDM to observations, most notably stellar kinematics in galaxies. Several groups have attempted to characterize the soliton-halo relation via numerical simulations. 

We performed a series of flat-space numerical simulations of ULDM halos initialized in different ways, with the aim of analyzing on the soliton-halo relation, exploring a larger range of $M_{\rm sol}/M$ than previous works. Our main findings are summarized in Figs.~\ref{fig:E_M} and \ref{fig:generic_final_snapshot}. In all of our simulations, solitons form around the $1/2$ relation, and then accrete mass from the surrounding halo without surpassing the $1/3$ relation. 
This can be explained by the $1/3$ relation being an upper bound, whereas the $1/2$ relation scales parametrically as, and is a factor of about 3 above the evaporation lower bound derived in Ref.~\cite{Chan:2022bkz}. 

After accounting for an ambiguity in the definition of the potential energy in a finite volume, our results are compatible with data from other works in the literature. 
We believe it would be useful if other groups who have carried out flat-space simulations re-analyzed their data taking into account this issue with the potential energy to check whether they also find solitons lying within the band between the $1/2$ and $1/3$ relations. 
If this is indeed the case, then the soliton-halo relation could be viewed as a ``soliton-halo band'' bracketed by these two relations. 

We have also noted that the quantity $\Xi
$ (defined in Eq.~(\ref{eq:Xi})), which is often used in the literature to express a soliton-halo relation, is sensitive to finite-volume effects.

If solitons still fall into the soliton-halo band once cosmological effects are included (as Fig.~\ref{fig:Chan_plot} suggests might be the case), then the use of the $1/2$ relation when investigating observational consequences can be regarded as reasonably conservative. In particular, we predict solitons to have mass roughly given by the $1/2$ relation or larger and as a result the inner peak in the  rotation curve, induced by the soliton, is generically of similar amplitude or larger than the outer one induced by the halo. 
An important remaining question is the timescale on which the soliton forms cosmologically: If the relaxation timescale is relevant then the soliton-halo predictions hold for dwarf galaxy systems (velocity dispersion of the order of 10~km/s) for ULDM masses $m \lesssim 10^{-20} \, \unit{\electronvolt}$. There could however be other faster timescales relevant in cosmological scenarios compared to relaxation (e.g. associated to the halo merger history). These might extend the applicability to the soliton-halo relation to larger ULDM particle masses. A detailed study of these aspects is left for future work. 
Another important question involves the role of baryon physics in shaping the soliton-halo relation. Although works on baryon feedback on ULDM already exist~\cite{Veltmaat:2019hou,Mocz:2019pyf}, it would be interesting to include such effects in our simulations in future work.

Solitons can also form inside small-scale overdensities in theories in which the Dark Matter is in the wave-like regime, $m\lesssim {\rm eV}$ but not ultra-light with $m\gg 10^{-20}\,{\rm eV}$, for example in theories of post-inflationary axions (see e.g. \cite{Kolb:1993zz,Eggemeier:2019jsu,Gorghetto:2024vnp}) and dark photons (see e.g.~\cite{Gorghetto:2022sue,Amin:2022pzv}), and it might be interesting to analyze the impact of the soliton-halo band in this context. 


\acknowledgments
We thank S. Sibiryakov for discussions and providing valuable insights. We also thank J. M. Camalich, P. Mocz, S. Sibiryakov and W. Xue for feedback on a draft of this work. KB and LT acknowledge the support by the European Research Area (ERA) via the UNDARK project (project number 101159929). The work of MG
is supported by the Alexander von Humboldt foundation and has been partially funded by the
Deutsche Forschungsgemeinschaft under Germany’s Excellence Strategy - EXC 2121 Quantum
Universe - 390833306. 
EH acknowledges the UK Science and Technology Facilities Council for support through the Quantum Sensors for the Hidden Sector collaboration under the grant ST/T006145/1 and UK Research and Innovation Future Leader Fellowship MR/V024566/1. For the purpose of open access, the author has applied a CC BY public copyright licence to any Author Accepted Manuscript (AAM) version arising from this submission.

\appendix
\section{Solving the Schr\"odinger--Poisson equations} \label{s:code}

The ULDM follows the Schr\"{o}dinger--Poisson equations for a non-relativistic field $ \psi $,
\begin{align} \label{eq:SPEfield}
	&\iu\pdv{\psi}{t} = -\frac{1}{2m} \laplacian{\psi} + m\Phi\psi \ , \\
    \label{eq:SPEfield1}
& \laplacian \Phi = 4\pi G(|\psi|^2 - \langle|\psi|^2|\rangle) \ ,		
\end{align}
where $\langle\cdot\rangle$ indicates the spatial average.  
These can be obtained from a relativistic scalar field 
\begin{equation}
\phi \equiv \frac{1}{\sqrt{2} m} (\psi \e^{-\iu m t} + \psi^* \e^{\iu m t} ) \ ,
\end{equation}
that obeys the Klein--Gordon equation, in the limit in which $\psi$ varies slowly with respect to $m$, i.e. $\partial_t\psi\ll m\psi$ and $\partial_t^2\psi \ll m^2\psi$. 
It is convenient to express the fields and coordinates in a dimensionless form. To this end, we define
\begin{equation} \label{eq:rescaling}
	\tilde{\psi} = \frac{1}{\lambda^2} \frac{\sqrt{4\pi G}}{m} \psi \ , \ \vec{\tilde{x}} = \lambda m \vec{x} \ , \ \tilde{t} = \lambda^2 m t \ , \
	\tilde{\Phi} = \Phi/\lambda^2 \ ,
\end{equation}
where $\lambda $ is any dimensionless parameter. With this definition, Eq.~\eqref{eq:SPEfield} becomes
\begin{align} \label{eq:SPtilde1}
	&\iu\pdv{\tilde{\psi}}{\tilde{t}} = -\frac{1}{2} \laplacian_{\tilde{x}}{\tilde{\psi}} + \tilde{\Phi} \tilde{\psi} \ , \\
	\label{eq:SPtilde2}
	& \laplacian_{\tilde{x}} \tilde{\Phi} = |\tilde\psi|^2   - \langle|\tilde\psi|^2|\rangle \ ,	
\end{align}
and is independent of $m$ and $G$, as well as of the rescaling parameter $\lambda$ (we omit the subscript in $\laplacian_{\tilde{x}}$ from now on).  

The density of the field is $\tilde{\rho}=|\tilde{\psi}|^2$, its total mass is 
\begin{equation}
	\tilde{M} = \int \dd[3]{\tilde{x}} |\tilde{\psi}|^2 \ ,
\end{equation}
and its energy is
\begin{equation}
	\tilde{E} =\tilde{E}_{\rm kin}+\tilde{E}_{\rm pot}= \int \dd[3]{\tilde{x}} \qty( \frac{|\grad{\tilde{\psi}}|^2}{2} + \frac{1}{2} |\tilde{\psi}|^2 \tilde{\Phi} ) \ .
\end{equation}
Both $M$ and $E$  are conserved under the Schr\"{o}dinger--Poisson evolution. Under the rescaling in Eq.~\eqref{eq:rescaling}, 
\begin{equation} \label{eq:tildeM}
\tilde{M} = \frac{4\pi G m}{\lambda} M \ ,  \ \ \tilde{E}
= \frac{4\pi G m}{\lambda^{3}} E
\ \implies \  \Xi \equiv \frac{|E|}{M^3} \frac{1}{G^2m^2} = \tilde{\Xi} \ .
\end{equation}
Thus, $\Xi$ is  invariant  
under the $\lambda$  rescaling.

We solve the rescaled Schr\"odinger--Poisson equations, in Eqs.~\eqref{eq:SPtilde1} and~\eqref{eq:SPtilde2}, using a standard pseudo-spectral 3D  
solver, similar to that described in Ref.~\cite{Levkov:2018kau}, in a periodic box with length $\tilde{L}$. In the following, we review the basics of its implementations, along with details specific to our setup. 
The field is evolved via the unitary operator
\begin{equation}
\tilde\psi(\tilde{t}+ \dd \tilde{t}) = \prod_{\alpha} \e^{-\iu d_\alpha \dd{\tilde{t}} \tilde{\Phi}_\alpha} \e^{-\iu c_\alpha\dd{\tilde{t}} \frac{(-\iu \grad)^2}{2} }\tilde\psi(\tilde{t}) \ ,
\end{equation}
where $\dd{\tilde{t}}$ is the time-step. In this equation, first the kinetic operator acts on $\tilde\psi(\tilde{t})$ as
\begin{equation}\label{eq:kin_op}
\e^{-\iu c_\alpha\dd{\tilde{t}} \frac{(-\iu \grad)^2}{2} }\tilde\psi(\tilde{t}) \equiv \tilde\psi^{(\alpha)}(\tilde{t} +\dd{\tilde{t}} ) \ ,
\end{equation}
followed by the potential operator, which is obtained by solving:
\begin{equation}\label{eq:Poiss_num}
\laplacian{\tilde\Phi_\alpha} = |\tilde\psi^{(\alpha)}|^2 -  \langle|\tilde\psi^{(\alpha)}|^2\rangle \ .
\end{equation}
The range that $\alpha$ runs over, and the values of the constants $ c_\alpha $, $ d_\alpha $, depend on the order of the numerical integrator that is used. We make use of the 6th order version, for which the numerical values of $ c_\alpha $, $ d_\alpha $ can be found in~\cite{Levkov:2018kau}.
We evaluate Eqs.~(\ref{eq:kin_op}) and (\ref{eq:Poiss_num}) in Fourier space,  via fast Fourier transforms using the FFTW library~\cite{10.1145/301631.301661}. 
We use adaptive time-steps, to ensure a conservation of energy $ \Delta E/E \lesssim 10^{-5} $ between time-steps, and an overall conservation of energy $ |E^{\rm tot}_{\rm final} - E^{\rm tot}_{\rm initial}|/|E^{\rm tot}_{\rm final} + E^{\rm tot}_{\rm initial}| \lesssim 10^{-3} $.

In particular, Eq.~(\ref{eq:Poiss_num}) is solved by setting the zero-momentum mode ($ k=0$) of $\tilde{\Phi}$ to zero, implying that the average of $\tilde\Phi $ over the box is zero, $ \langle{\tilde\Phi}\rangle =0 $. 
To fix the arbitrary constant in the potential energy $E_{\rm pot}$, we define
\begin{equation} \label{eq:c}
	\tilde\Phi(\tilde{x}) \equiv \tilde\Phi_{\rm code}(\tilde{x}) +c \ ,
\end{equation}
where $ \tilde\Phi_{\rm code} $ is the gravitational potential computed as described above and satisfying $\langle\tilde\Phi_{\rm code}\rangle=0$. Via an appropriate determination of the constant $ c $, we then require $\lim_{\tilde{r}\to \infty} \tilde{\Phi}(\tilde{r}) =0$, where $\tilde{r}$ is the distance to the point of maximum density. Specifically, from the Poisson equation, 
the asymptotic behavior of $ \tilde\Phi  $ in the infinite-volume limit is
\begin{equation} \label{eq:Phi_big_r}
	\tilde\Phi \sim -\frac{\tilde{M}}{4\pi \tilde{r}} \text{ for large $ \tilde{r} $} \ . 
\end{equation}
Hence, if $ \tilde{R} $ is the furthest point in the grid from the point of maximum density (within which we interpret all the mass $ \tilde{M} $ of the box to lie), we can approximately recover the constant $ c $ as
\begin{equation} \label{eq:phi_const}
	\tilde\Phi(\tilde{R}) = \tilde\Phi_{\rm code}(\tilde{R}) +c = -\frac{\tilde{M}}{4\pi \tilde{R}} \implies c = -\frac{\tilde{M}}{4\pi\tilde{R}} - \Phi_{\rm code}(\tilde{R}) \ .
\end{equation} 
We stress that this method is just a rough estimate, since there is no guarantee that the potential already goes as Eq.~(\ref{eq:Phi_big_r}) within the box.

\begin{figure}
    \centering
    \includegraphics[width=0.7\linewidth]{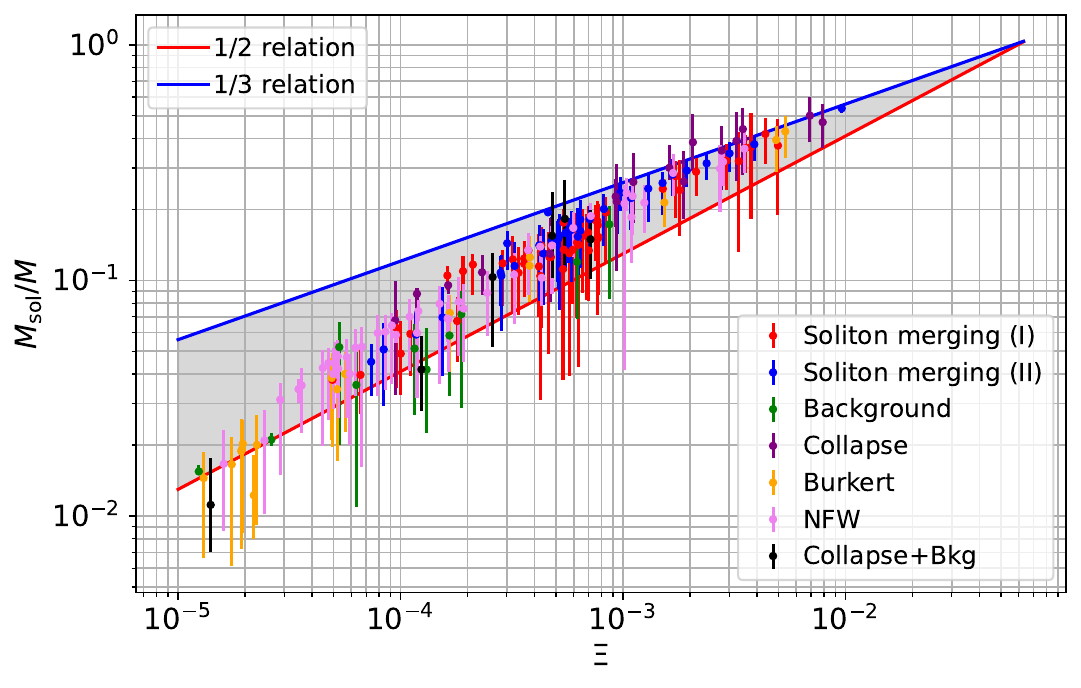}
    \caption{Same as in Fig.~\ref{fig:generic_final_snapshot}, but with $\Xi$ calculated using $E_{\rm tot}$ instead of $E_{\rm kin}$. Unlike Fig.~\ref{fig:etot_check} left, the constant in the potential energy (which in enters $\Xi$) is chosen as in Eq.~\eqref{eq:phi_const} such this matches its infinite-volume value.}
    \label{fig:etot_plot}
\end{figure}

In Fig.~\ref{fig:etot_plot}, we plot our soliton-halo data points as a function of $\Xi$, calculated from the total energy $E_{\rm tot}$ including the constant Eq.~(\ref{eq:phi_const}) in  $E_{\rm pot}$ (with $\tilde{R}=\tilde{L}/2$). Relative to Fig.~\ref{fig:generic_final_snapshot} (which is the same plot but with $\Xi$ calculated from the kinetic energy alone), many of the data points are shifted to the left. We believe this discrepancy is likely due to the calculation of the constant in the potential energy via Eq.~(\ref{eq:phi_const}) not being accurate enough. On the other hand, with the corrected $E_{\rm tot}$ the points are  substantially closer to the results using $E_{\rm kin}$ than when using the uncorrected $E_{\rm tot}$, as plotted in Fig.~\ref{fig:etot_check}. This suggests that our calculation of the constant in the potential energy is at least partially successful. Given this situation, and the fact that a mis-determination of this constant can affect the interpretation of results, we decided to use quantities calculated from $E_{\rm kin}$ for our main plots.

As mentioned in Section~\ref{s:results}, we discard a few runs that do not approximately satisfy the virial theorm (i.e. for which the system is not virialized); in particular, we demand that
\begin{equation} \label{eq:virial_criterium}
(2E_{\rm kin} -| E_{\rm pot}|)/(2E_{\rm kin} +| E_{\rm pot}|) < 0.2 \ .
\end{equation}

\section{Soliton solution} \label{s:soliton}
The soliton solution can be obtained from Eqs.~(\ref{eq:SPtilde1}) and (\ref{eq:SPtilde2}) with the spherically-symmetric ansatz
\begin{equation}
	\tilde{\psi} (\tilde{\vec{x}}, \tilde{t}) \equiv \chi(\tilde{r}) \e^{-\iu \tilde\omega_{\rm sol} \tilde{t}} \ ,
\end{equation}
which leads to
\begin{align} \label{eq:Sol1}
	&\partial^2_{\tilde{r}}(\tilde{r} \chi)= 2\tilde{r}(\tilde{\Phi} - \tilde{\omega}_{\rm sol}) \chi \ , \\
	\label{eq:Sol2}
	& \partial^2_{\tilde{r}}(\tilde{r} \tilde\Phi)= \tilde{r} \chi^2 \ .	
\end{align}
These equations are invariant under the $ \lambda $ rescaling of Eq.~(\ref{eq:rescaling}) and thus admit an infinite class of solutions for $\chi$
, labeled by $ \chi_\lambda $. We define them such that $ \chi_\lambda(0) = \lambda^2 $. We can solve the equations above for $ \chi_1(0)$ with boundary conditions $\partial_{\tilde r}\chi(0)=\partial_{\tilde r}\Phi(0)=0$, $\lim_{\tilde{r}\to \infty}\chi(\tilde r)=0$ 
to obtain
\begin{equation}
\tilde{M}_{\rm sol,1} \approx 26 \ , \ \tilde{r}_{\rm c, 1} \approx 1.31 \ , \ \tilde\omega_{\rm sol,1} \approx -0.69 + \tilde\Phi(\tilde{r}\to \infty) \ .
\end{equation}
Here, $\tilde{M}_{\rm sol}=\int d^3\tilde{x}|\tilde{\psi}|^2=\int d^3\tilde{x}{\chi}^2$ is the total soliton mass and $\tilde{r}_{\rm c}$ is the soliton half-density radius, i.e. $|\tilde{\psi}(\tilde{r}_{\rm c})|^2=|\tilde{\psi}(0)|^2/2$. For an isolated soliton, $ \tilde\Phi(\tilde{r}\to \infty)=0$.\footnote{For a soliton surrounded by a halo, one can identify $ \tilde\Phi(\tilde{r}\to \infty) $ with $ \tilde{\Phi}_{\rm offset} $ in Eq.~(\ref{eq:phi_offset}).} A generic soliton will then have
\begin{equation} \label{eq:lambda_resc}
\tilde{M}_{\rm sol} = \lambda \tilde{M}_{\rm sol,1}  \ , \  \tilde{E}_{\rm sol}  = \lambda^3\tilde{E}_{\rm sol,1} \ , \  \ \tilde{r}_{\rm c}  = \tilde{r}_{\rm c,1}/\lambda \ , \ \ \tilde\omega_{\rm sol} \approx -0.69\lambda^2 + \tilde\Phi(\tilde{r}\to \infty) \ .
\end{equation} 
From these equations it follows $r_{\rm c}=1.31/(4\pi Gm^2\rho(0))^{1/4}$ and $ M^4_{\rm sol} = 64\pi\rho(0) / (G^3 m^6)$, used in Section~\ref{s:sim}. 

The soliton is a virialized system, i.e. $ 2\tilde{E}_{\rm kin,sol} = -\tilde{E}_{\rm pot,sol} $, and one can show that
\begin{equation} \label{eq:sol_etot}
\tilde{E}_{\rm sol} = \frac{1}{3} \tilde{M}_{\rm sol} \tilde{\omega}_{\rm sol} = -\tilde{E}_{\rm kin,sol}\  ,
\end{equation}
where in the first equality we integrated by parts the equations of motions and used the virial theorem. Hence, since $ \tilde{E}_{\rm sol}/\tilde{M}_{\rm sol}^3 =  \tilde{E}_{{\rm sol},1}/\tilde{M}_{\rm sol,1}^3$ given that $\Xi$ is independent of $\lambda$, a soliton-halo relation of the form in Eq.~(\ref{eq:sol_halo_rel}) whose validity is retained in the limit $M_{\rm sol}/M \to 1$ should always obey
\begin{equation} \label{eq:single_sol}
1 = \alpha\left(\frac{|E_{\rm sol,1}|}{G^2m^2M_{\rm sol,1}^3}\right)^\beta=\alpha \qty((4\pi)^2 \frac{\tilde{\omega}_{\rm sol,1}}{3\tilde{M}^2_{\rm sol,1}} )^\beta \approx \alpha (0.054)^\beta \ . 
\end{equation}
This implies that for a soliton one has $\Xi \approx 0.054$. 
Both $ \alpha = 4.2 $, $ \beta=1/2 $ (from~\cite{Schive:2014hza}) and $ \alpha=2.6 $, $\beta= 1/3 $ (from~\cite{Mocz:2017wlg}) approximately satisfy Eq.~(\ref{eq:single_sol}).  
From this result, one can infer that the $1/2$ relation is simply rewritten as in Eq.~(\ref{EMEM}), whereas the $1/3$ relation as in Eq.~(\ref{eq:13_simple}).

\section{More details on the initial conditions} \label{s:initial_cond}
In this Appendix, we present further details on the 
initial conditions of our simulations described in Section~\ref{ss:init}.

\paragraph{Halos.} 
We initialize halos via the Eddington procedure (see~\cite{Widrow:1993qq,Lancaster:2019mde})
\begin{equation}
	\psi(x) = (\Delta v)^{3/2}\sum_{\vec{v}} \sqrt{f(\mathcal{E}(x,v))} \e^{\iu m\vec{x}\cdot\vec{v} + \iu \varphi_{\vec{v}}} \ ,
\end{equation}
where $ \varphi_{\vec{v}} $ is a random phase dependent on $ \vec{v} $, $ \Delta v $ is the velocity spacing allowed by resolution in the simulation, and 
\begin{align}
	&f(\mathcal{E}) = \frac{2}{\sqrt{8}\pi^2} \int_0^{\sqrt{\mathcal{E}}} \dd{Q} \dv{^2\rho}{\Psi^2}{(Q)} \ , \\ 
	&\mathcal{E} = \Psi(r) - \frac{v^2}{2}  \ , \ \Psi = -\Phi + \Phi(r_{\rm max})  \ ,
\end{align}
where $  Q = \sqrt{\mathcal{E}-\Psi} $ and $ \Phi $ is the overall gravitational potential. As target density, we use a NFW profile~\cite{Navarro:1996gj} or a Burkert profile~\cite{Salucci:2000ps,Burkert:2015vla},
\begin{equation} \label{eq:nfw_burk}
\rho_{\rm nfw} = \frac{\rho_0 r^3_{\rm s}}{r(r+r_{\rm s})^2} \ , \ \rho_{\rm burk} = \frac{\rho_0 r^3_{\rm s}}{(r+r_{\rm s})(r^2+r_{\rm s}^2)} \ ,
\end{equation}
with varying $ \rho_0 $ and $ r_{\rm s} $.

\paragraph{Background.} 
We can satisfy Eq.~(\ref{eq:cov}) via\footnote{The field in Eq.~(\ref{eq:psi_noise}) is equivalent to the initialization in~\cite{Levkov:2018kau}, which reads 
$	\tilde\psi_{\tilde{k}} = \sqrt{8\tilde{N}} \pi^{3/4} \e^{-\tilde{k}^2/2} \e^{\iu \varphi_{\tilde{k}}} \ , $
where $ \varphi_{\tilde{k}} $ is uniformly distributed between $ 0 $ and $ 2\pi $, 
with the identifications $\bar{\rho} = \tilde{N} /L^3$ and $k_{\rm bkg} =1 \ $.
}
\begin{equation}\label{eq:psi_noise}
	\psi(x) = \frac{(\Delta k)^3}{L^3} \sum_{\vec{k}} \sqrt{\frac{\hat{C}_{\vec{k}}L^3}{2}} (A_{\vec{k}}  + \iu B_{\vec{k}}) \e^{\iu\vec{k}\cdot\vec{x}}  \ ,  \ \ \ \  \hat{C}_{\vec{k}} = \frac{8\bar{\rho} \pi^{3/2}}{k^3_{\rm bkg}} \e^{- k^2/k^2_{\rm bkg}} \ .
\end{equation}
Here, $ L^3 $ is the volume of the integration (i.e. the simulation volume), $ \Delta k =2\pi/L$ is the resolution in Fourier space
, and $ A_{\vec{k}} $ and $B_{\vec{k}}$ are uncorrelated Gaussianly-distributed variables with unitary variance and zero mean, satisfying
\begin{equation}
\expval{A_{\vec{k}}A_{\vec{k}'}} =  \expval{B_{\vec{k}}B_{\vec{k}'}} = \delta_{\vec{k}\vec{k}'} \ , \ \expval{A_{\vec{k}}B_{\vec{k}'}} = 0 \ .
\end{equation}


The kinetic energy per unit volume is
\begin{equation} \label{eq:kgas}
	\frac{E_{\rm kin}}{L^2}=\frac{1}{2m^2}\expval{|\grad{\psi}|^2} = \frac{1}{2m^2}\int \frac{\dd[3]{k}}{(2\pi)^3} \hat{C}_k k^2 =\frac{1}{2m^2} \frac{3 \bar{\rho} k^2_{\rm bkg}}{2}=\frac{3k^2_{\rm bkg}}{4} \frac{M}{L^3} \ , 
\end{equation}
where $M=\bar{\rho}L^3$. From this, it follows that $E_{\rm kin}/M=3k_{\rm bkg}^2/4m^2$.\footnote{This differs by a factor of 2 from the more common $\frac{E_{\rm kin}}{M}=\frac32 \frac{k_{\rm bkg}^2}{m^2}$ because the distribution has variance $k_{\rm bkg}^2/2$ rather than $k_{\rm bkg}^2$.}

\section{The criterion for soliton formation}\label{s:sol_form_criterium}

Here, we provide details of the criterion used to define the formation time of a soliton–halo system, as briefly discussed in Section~\ref{ss:when}. Explicitly:
\begin{enumerate}
    \item At each time step in the simulation we identify the point in the box that contains the highest mass density. Call this point the origin of coordinates $r=0$, with density $\rho(0)$. 
    Using the soliton profile approximation~\cite{Schive:2014dra} 
\begin{equation} \label{eq:fit}
\rho_{\rm fit}(r) = \frac{\lambda^4}{(1+a^2\lambda^2r^2)^b} \ , \ \ a\approx 0.228 \ \ , \ \ b\approx 4.071 \ , 
\end{equation}
we obtain a first estimate of the soliton core radius $ r_{\rm c}$ from the value of $\rho(0)$.
\item Next, we compute the radial-averaged density profile around $r=0$. We fit a soliton profile to the grid region surrounding $r=0$, treating $r_{\rm c}$ as a fit parameter, and including in the fit grid points within $r<3r_{\rm c}$. 

At the same time, to constrain the halo, we also fit an NFW profile to grid points satisfying $r>4 r_{\rm c}$, with the NFW scale radius $r_{\rm s}$ and density $\rho_{0}$ as fit parameters. Namely, we add
\begin{equation}
    \rho_{\rm fit}(r>4 r_{\rm c})=\frac{\rho_{0}}{\frac{r}{r_{\rm s}}\left(1+\frac{r}{r_{\rm s}}\right)^2}\ .
\end{equation}
\item The earliest time $t_{\rm form}$ that satisfies
\begin{equation} \label{eq:good_fit}
\frac{1}{N}\sum_{i\in\text{fit points}}^{N} \log^2\qty(\frac{\rho_{\rm sim}(r_i,t_{\rm form})}{\rho_{\rm fit}(r_i)}) < 0.5 \ ,
\end{equation}
is where we begin to report results in the context of a soliton-halo relation. 
\end{enumerate}

\begin{figure}
    \centering
\includegraphics[width=0.49\linewidth]{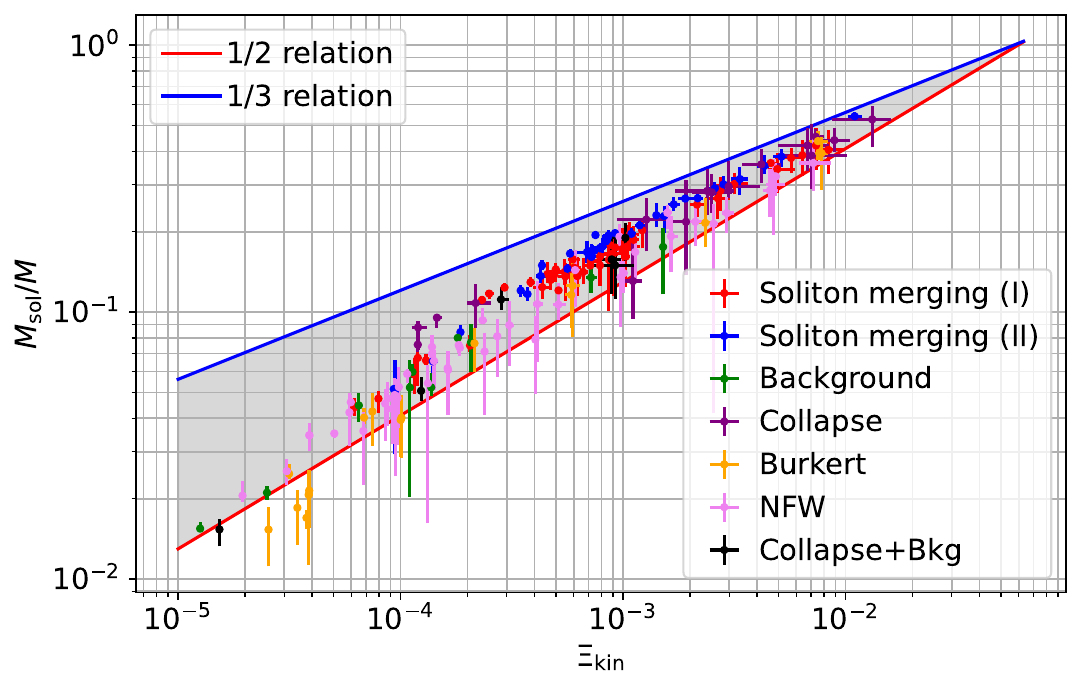}
    \includegraphics[width=0.49\linewidth]{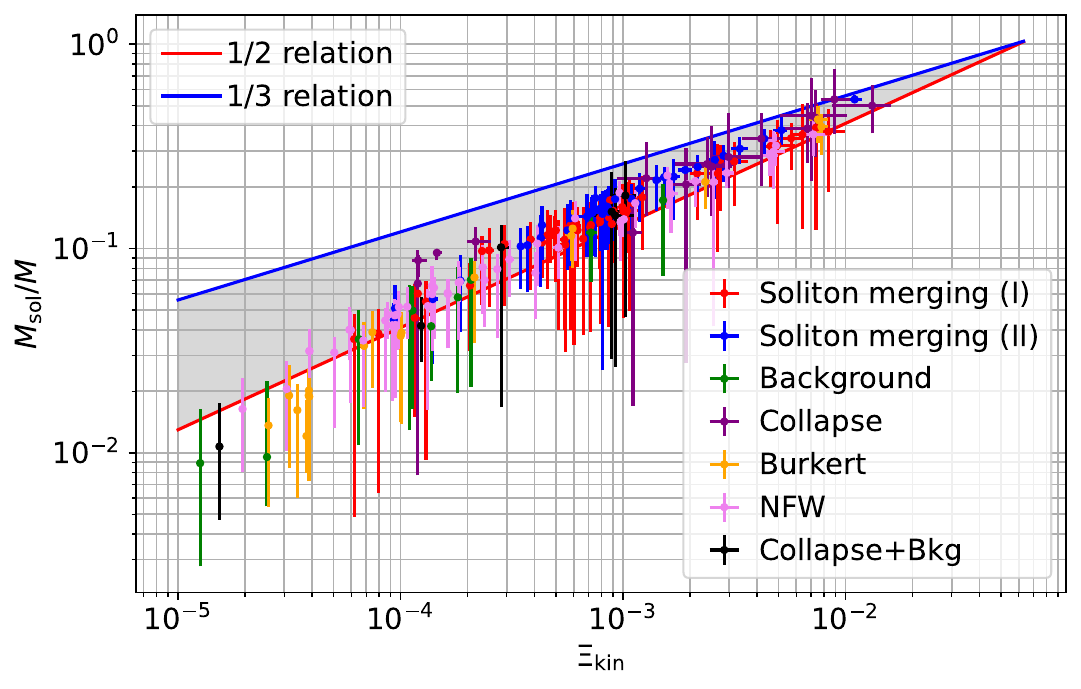}
    \caption{The same plot as in Fig.~\ref{fig:generic_final_snapshot}, but with a different thresholds used Eq.~(\ref{eq:good_fit}) for determining the moment of soliton formation: $0.05$ and $5$ in left and right panel respectively, instead of $0.5$.}
    \label{fig:check_sol_form}
\end{figure}

As we show in Fig.~\ref{fig:check_sol_form}, changing the threshold in Eq.~(\ref{eq:good_fit}) does not significantly change the results we obtain. A more restrictive criterion selects later formation times, making points in the soliton-halo diagram start their life when they are more massive. Meanwhile, a less restrictive criterion, e.g. increasing the threshold in Eq.~(\ref{eq:good_fit}) to $5$, initially less massive solitons. However, such a criterion is actually so weak that in some runs an examination of the field and density profile suggests there is actually no soliton at some times when the threshold is met, limiting the reliability of the soliton-halo relation plot obtained.

\section{The energy spectrum} \label{s:spectrum}
To gain deeper insight into soliton formation, one useful tool is the energy spectrum, defined as
\begin{equation}
F(t, \omega) = \dv{N}{\omega} \ .
\end{equation}
This is the number $ N $ of ULDM particles with a certain (non-relativistic) energy $ \omega $. As the soliton forms, we expect $ F(t,\omega_{\rm sol}) $ to increase accordingly. We briefly review relevant features of the energy spectrum.

$ F(t, \omega) $ was originally defined in~\cite{Levkov:2018kau} as
\begin{equation}
	F(t, \omega) = \int_{-\infty}^{\infty} \frac{\dd{t'}}{2\pi} \int \dd[3]{x} \psi^*(\vec{x},t) \psi(\vec{x},t') \e^{\iu \omega (t'-t) - (t-t')^2/\tau^2} \ ,
\end{equation}
where $ \Delta \omega\,\tau \gg 1 $, with $ \Delta \omega $ the energy resolution and $\tau$ a parameter of the Gaussian kernel in the integrand. This expression is justified by considering $\psi$ as a superposition of orthonormal energy eigenmodes $\psi_n$:
\begin{equation} \label{eigen}
	\psi = \sum_n c_n(t) \psi_n(\vec{x}) \e^{-\iu \omega_n t} \ .
\end{equation} 
Using $ \int\dd[3]{x} \psi^*_n(\vec{x})\psi_{n'}(\vec{x}) = \delta_{nn'} $, we have, with $ t'-t \equiv \Delta t$,
\begin{equation}
	F(t,\omega) = \sum_n c^*_n(t)  \int \frac{\dd{\Delta t}}{2\pi} c_n(t + \Delta t)  \e^{\iu \Delta t (\omega_n - \omega)  - \Delta t^2/\tau^2} \ ;
\end{equation}
if we assume that in the integration range $ c_n(t + \Delta t) \approx c_n(t) $, we can complete the square inside the integral to recover a Gaussian integral and obtain
\begin{equation} \label{Fresult}
	F(t,\omega) = \sum_n |c_n(t)|^2 \frac{\tau}{2\sqrt{\pi}} \e^{-(\omega - \omega_n)^2\tau^2/4} \approx  \sum_n |c_n(t)|^2 \delta(\omega_n -\omega) \ .
\end{equation} 
Interestingly, the same result can be derived using an alternative, manifestly real, formula:
\begin{equation} \label{Fnew}
	F(t,\omega) = 2 \Re  \int_0^\infty \frac{\dd{t'}}{2\pi} \int \dd[3]{x} \psi^*(\vec{x},t) \psi(\vec{x},t+t') \e^{\iu \omega t' - {t'}^2/\tau^2} \ .
\end{equation} 
One can again plug the ansatz of Eq.~(\ref{eigen}) and see that Eq.~(\ref{Fresult}) holds. We can also write the previous expression as
\begin{equation} \label{F_alt}
	F(t,\omega) =  \int_{-\infty}^\infty \frac{\dd{t'}}{2\pi} \int \dd[3]{x} \psi^*(\vec{x},t_1) \psi(\vec{x},t_1+t') \e^{\iu \omega t' - {t'}^2/\tau^2} \ , \ t_1 = 
	\begin{cases}
		t - t' &\text{ for } t' < 0 \ , \\
		t &\text{ for } t' \ge 0 \ . 
	\end{cases}
\end{equation}
The formulation Eq.~(\ref{Fnew}) is more convenient, since in a simulation one can compute Eq.~(\ref{Fnew}) by storing only one snapshot at time $ t $ and perform the integral until the next snapshot; this avoids using a large amount of memory storing multiple snapshots.

In the grid, assuming equal time-steps with spacing $ \Delta t $, we have
\begin{equation} \label{tildeF}
	\tilde{F} =  \sum_n |c_n(\tilde{t})|^2 \frac{\Delta \tilde{t}}{\pi}  \sum_{l=0} \e^{-(l\Delta \tilde{t})^2/\tilde\tau^2} \cos(l\Delta \tilde{t} (\tilde{\omega} - \tilde{\omega}_n)) \ , \ \tilde{\omega} = \frac{\omega}{\lambda^2 m} \ , \ \tilde{F} = 4\pi G m \lambda F \ .
\end{equation}
In particular, this function is periodic under
\begin{equation}
	\tilde{\omega} \to \tilde{\omega} + \frac{2\pi p}{\Delta \tilde{t}} \ , \ p \in \mathbb{Z} \ .
\end{equation}
The numerically computed $ \tilde{F} $ then makes sense only in a range smaller than the period $ 2\pi/\Delta \tilde{t} $.
For Eq.~(\ref{tildeF}) to reduce to 
\begin{equation}
	\tilde{F} \to \frac{1}{\pi} \int_0^\infty \dd{t'} \e^{-(t')^2/\tilde\tau^2} \cos(t' (\tilde{\omega} - \tilde{\omega}_n)) = \frac{1}{\sqrt{2\pi}\sigma} \e^{\frac{(\tilde{\omega} - \tilde{\omega}_n)^2}{2\sigma^2} } \ , \ \sigma \equiv \frac{\sqrt{2}}{\tau} \ , 
\end{equation}
we need to impose
\begin{equation}
	\Delta \tilde{\omega} \tau \gg 1 \ , \ \Delta \tilde{\omega} \Delta \tilde{t} \ll 1 \ , \ \frac{\tau}{\Delta \tilde{T}} \ll 1 \ ,
\end{equation}
where $ \Delta \tilde{T} $ is the time between snapshots.

The dimensionless $ c_n $ coefficients relate to the number of particles in state $ n $,
\begin{equation}
	\tilde{F} = \dv{\tilde{N}}{\tilde{\omega}} \sim |c_n|^2 \delta(\omega - \tilde{\omega}_n) \ .
\end{equation}
One can see that, from Eq.~(\ref{F_alt}), we have
\begin{equation}
	\int \dd{\tilde{\omega}} \tilde{F} = \tilde{M} \ , \ \ \ \  \tilde{E}_{\rm part} \equiv \int \dd{\tilde{\omega}} \tilde{\omega} \tilde{F} \ ,
\end{equation}
where $E_{\rm part}$ is the  sum of the energy of all the particles in the box. 
Working out the last equality, one finds out that $ \tilde{F} $ encodes the particle energy, where the energy is the sum of the kinetic energy and the contribution of the gravitational potential, the latter viewed as an external potential. In particular, $ \tilde{E}_{\rm part} \neq \tilde{E}_{\rm tot} $, where $ \tilde{E}_{\rm tot} $ is the total energy of the system, where the fact that the gravitational potential is self-sourced is taken into account. In fact,
\begin{equation}
	\int \dd{\omega} \omega \e^{\iu \omega t'} = -2\pi\iu \pdv{t'} \delta(t') \implies \int \dd{\omega} \omega F = \int \dd[3]{x} \psi^* \iu \pdv{\psi}{t} = \int \dd[3]{x} \frac{|\grad{\psi}|^2}{2} + \int \dd[3]{x}\Phi |\psi|^2 \ , 
\end{equation}
without a factor of $ 1/2 $ on the potential energy contribution.

Notice that $ \tilde\omega \sim \tilde{E}_{\rm kin} + \tilde\Phi $, in particular a constant shift in $ \tilde\Phi $ would yield a constant shift in $ \tilde\omega $; from Eq.~(\ref{eq:phi_const}), we can write
\begin{equation}
	\tilde{F}_{\rm code} (\tilde\omega_{\rm code}) =  \tilde{F} (\tilde\omega) \ , \ \ \ \ \tilde\omega = \tilde\omega_{\rm code} + c \ ,
\end{equation} 
where $\tilde{F}_{\rm code}$, $\tilde\omega_{\rm code}$ are the quantities coming from our Schr\"odinger--Poisson solver, and $c$ is the constant in Eq.~\eqref{eq:c}.
When a halo is present, $ \omega_{\rm sol} $ gets shifted due to the further potential well coming from the halo. To understand the origin of such a shift, one can look at the gravitational potential for a spherical distribution:
\begin{equation}
	\Phi(r) = - \frac{G(M_{\rm sol}(r) + M_{\rm halo}(r)) }{r} - 4\pi G \int_r^\infty \dd{y} y (\rho_{\rm sol}(y) + \rho_{\rm halo}(y)) \ .
\end{equation}
We further approximate $ \rho_{\rm halo}(r) \approx 0  $ for $ r < r_{\rm t} $ and $ \rho_{\rm sol}(r) \approx 0 $ for $ r < r_{\rm t} $, where $ r_{\rm t} $ is a transition radius (which can be taken to be $ 3 r_{\rm c} $). Then, for $ r<r_{\rm t} $,
\begin{equation} \label{eq:phi_offset}
	\Phi(r) = \Phi_{\rm sol}(r) - 4\pi G \int_{r_{\rm t}}^\infty \dd{y} y \rho_{\rm halo}(y) \equiv\Phi_{\rm sol}(r)  + \Phi_{\mathrm{offset}} \ ,
\end{equation} 
and hence
\begin{equation}
	\tilde\omega^{\rm sol}_{\rm code} + c = \tilde\omega_{\mathrm{no\,halo}}^{\rm sol} +  \tilde\Phi_{\mathrm{offset}} \ ,
\end{equation}
where again $\tilde\omega^{\rm sol}_{\rm code}$ is the location of the soliton energy as coming out from our Schr\"odinger--Poisson solver, while $\tilde\omega_{\mathrm{no\,halo}}^{\rm sol}$ is the actual energy corresponding to an isolated soliton. 

\begin{figure}
	\centering
	\includegraphics[width=0.4\textwidth]{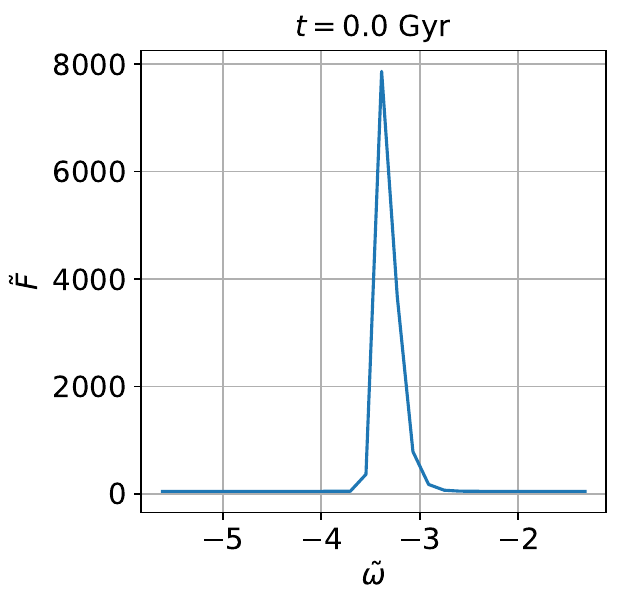}
	\includegraphics[width=0.4\textwidth]{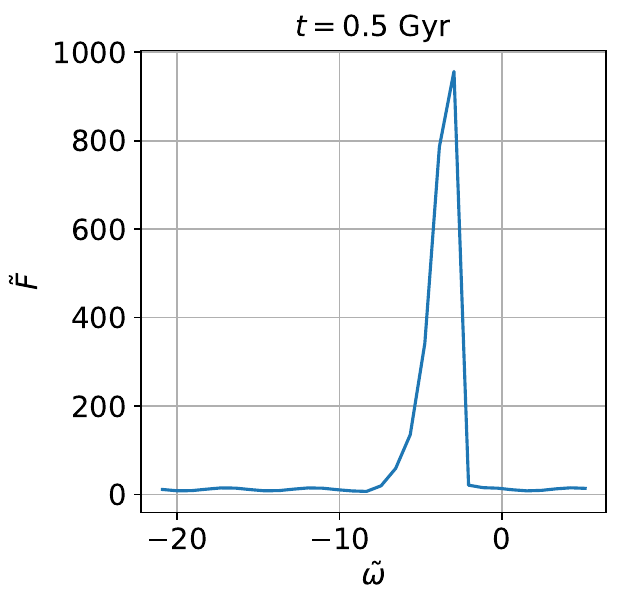}
	\includegraphics[width=0.4\textwidth]{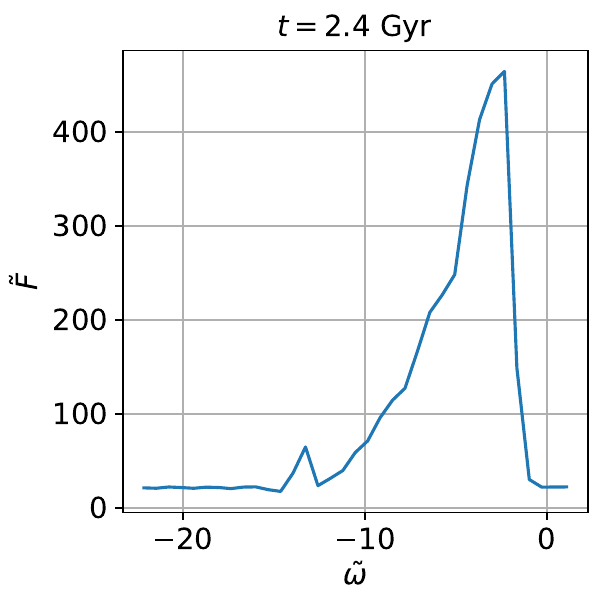}
\includegraphics[width=0.4\textwidth]{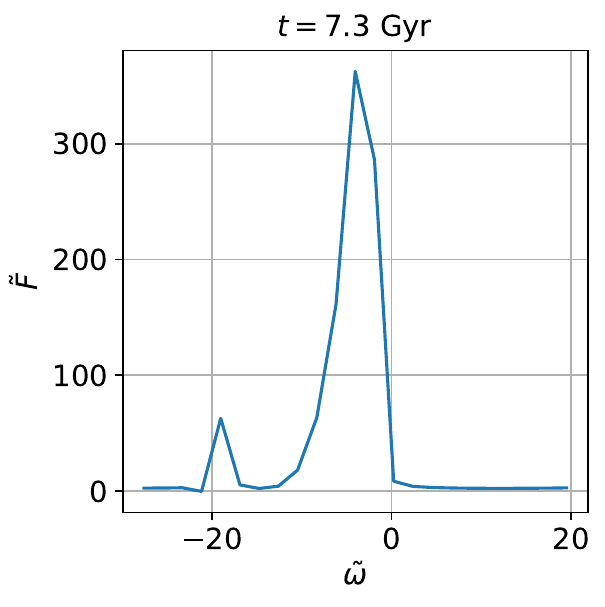}
	\caption{The spectrum $ \tilde{F}(\tilde{\omega}) $ for the same simulation and times as plotted in Fig.~\ref{fig:sample_run}. One can see a peak forming (corresponding to the soliton) and separate from the main peak corresponding to the halo. In particular, at $t=7.3\,{\rm Gyr}$ the peak lies at around $ \tilde{\omega} \approx -19 $; integrating the area in the vicinity of that peak, one finds $ \tilde{M}_{\rm sol} \approx 100 $, consistent with the soliton mass obtained by fitting of the radially-averaged density profile. Notice that an isolated soliton of that mass would have $ \tilde{\omega} \approx -10 $. The difference is due to $ \Phi_{\mathrm{offset}} $, Eq.~(\ref{eq:phi_offset}).}
	\label{fig:F_snap}
\end{figure}

In Fig.~\ref{fig:F_snap} we show an example of $ \tilde{F} $ for the snapshots of the simulation plotted in Fig.~\ref{fig:sample_run}. One can see the formation of the soliton via a peak arising at $ \tilde\omega_{\mathrm{no\,halo}}^{\rm sol} +  \tilde\Phi_{\mathrm{offset}}  $. The computation of $F(\omega,t)$ is an interesting cross-check of our main method adopted for defining the moment of soliton formation, Eq.~(\ref{eq:good_fit}). However, using $F(\omega,t)$ there remain uncertainties, in particular when to call the forming peak a soliton. These mean that this method of identifying the formation time does not have a clear advantage compared to simply considering the density profile.

\bibliography{ref}

\providecommand{\href}[2]{#2}\begingroup\raggedright\begin{thebibliography}{10}

\bibitem{Preskill:1982cy}
J.~Preskill, M.~B. Wise, and F.~Wilczek, ``{Cosmology of the Invisible Axion},'' \href{http://dx.doi.org/10.1016/0370-2693(83)90637-8}{{\em Phys. Lett. B} {\bfseries 120} (1983) 127--132}.

\bibitem{Abbott:1982af}
L.~F. Abbott and P.~Sikivie, ``{A Cosmological Bound on the Invisible Axion},'' \href{http://dx.doi.org/10.1016/0370-2693(83)90638-X}{{\em Phys. Lett. B} {\bfseries 120} (1983) 133--136}.

\bibitem{Dine:1982ah}
M.~Dine and W.~Fischler, ``{The Not So Harmless Axion},'' \href{http://dx.doi.org/10.1016/0370-2693(83)90639-1}{{\em Phys. Lett. B} {\bfseries 120} (1983) 137--141}.

\bibitem{Svrcek:2006yi}
P.~Svrcek and E.~Witten, ``{Axions In String Theory},'' \href{http://dx.doi.org/10.1088/1126-6708/2006/06/051}{{\em JHEP} {\bfseries 06} (2006) 051},
\href{http://arxiv.org/abs/hep-th/0605206}{{\ttfamily arXiv:hep-th/0605206 [hep-th]}}.

\bibitem{Arvanitaki:2009fg}
A.~Arvanitaki, S.~Dimopoulos, S.~Dubovsky, N.~Kaloper, and J.~March-Russell, ``{String Axiverse},'' \href{http://dx.doi.org/10.1103/PhysRevD.81.123530}{{\em Phys. Rev.} {\bfseries D81} (2010) 123530},
\href{http://arxiv.org/abs/0905.4720}{{\ttfamily arXiv:0905.4720 [hep-th]}}.

\bibitem{Marsh:2015xka}
D.~J.~E. Marsh, ``{Axion Cosmology},'' \href{http://dx.doi.org/10.1016/j.physrep.2016.06.005}{{\em Phys. Rept.} {\bfseries 643} (2016) 1--79},
\href{http://arxiv.org/abs/1510.07633}{{\ttfamily arXiv:1510.07633 [astro-ph.CO]}}.

\bibitem{Hui2017}
L.~Hui, J.~P. Ostriker, S.~Tremaine, and E.~Witten, ``{Ultralight scalars as cosmological dark matter},'' \href{http://dx.doi.org/10.1103/PhysRevD.95.043541}{{\em Phys. Rev.} {\bfseries D95} no.~4, (2017) 043541},
\href{http://arxiv.org/abs/1610.08297}{{\ttfamily arXiv:1610.08297 [astro-ph.CO]}}.

\bibitem{Lague:2021frh}
A.~Lagu\"e, J.~R. Bond, R.~Hlo\v{z}ek, K.~K. Rogers, D.~J.~E. Marsh, and D.~Grin, ``{Constraining ultralight axions with galaxy surveys},'' \href{http://dx.doi.org/10.1088/1475-7516/2022/01/049}{{\em JCAP} {\bfseries 01} no.~01, (2022) 049}, \href{http://arxiv.org/abs/2104.07802}{{\ttfamily arXiv:2104.07802 [astro-ph.CO]}}.

\bibitem{Irsic:2017yje}
V.~Ir\v{s}i\v{c}, M.~Viel, M.~G. Haehnelt, J.~S. Bolton, and G.~D. Becker, ``{First constraints on fuzzy dark matter from Lyman-$\alpha$ forest data and hydrodynamical simulations},'' \href{http://dx.doi.org/10.1103/PhysRevLett.119.031302}{{\em Phys. Rev. Lett.} {\bfseries 119} no.~3, (2017) 031302},
\href{http://arxiv.org/abs/1703.04683}{{\ttfamily arXiv:1703.04683 [astro-ph.CO]}}.

\bibitem{Armengaud:2017nkf}
E.~Armengaud, N.~Palanque-Delabrouille, D.~J.~E. Marsh, J.~Baur, and C.~Yï¿œche, ``{Constraining the mass of light bosonic dark matter using SDSS Lyman-$\alpha$ forest},'' \href{http://dx.doi.org/10.1093/mnras/stx1870}{{\em Mon. Not. Roy. Astron. Soc.} {\bfseries 471} no.~4, (2017) 4606--4614},
\href{http://arxiv.org/abs/1703.09126}{{\ttfamily arXiv:1703.09126 [astro-ph.CO]}}.

\bibitem{Kobayashi:2017jcf}
T.~Kobayashi, R.~Murgia, A.~De~Simone, V.~Ir\v{s}i\v{c}, and M.~Viel, ``{Lyman-$\alpha$ constraints on ultralight scalar dark matter: Implications for the early and late universe},'' \href{http://dx.doi.org/10.1103/PhysRevD.96.123514}{{\em Phys. Rev. D} {\bfseries 96} no.~12, (2017) 123514}, \href{http://arxiv.org/abs/1708.00015}{{\ttfamily arXiv:1708.00015 [astro-ph.CO]}}.

\bibitem{Leong:2018opi}
K.-H. Leong, H.-Y. Schive, U.-H. Zhang, and T.~Chiueh, ``{Testing extreme-axion wave dark matter using the BOSS Lyman-Alpha forest data},'' \href{http://dx.doi.org/10.1093/mnras/stz271}{{\em Mon. Not. Roy. Astron. Soc.} {\bfseries 484} (2019) 4273},
\href{http://arxiv.org/abs/1810.05930}{{\ttfamily arXiv:1810.05930 [astro-ph.CO]}}.

\bibitem{Lancaster:2019mde}
L.~Lancaster, C.~Giovanetti, P.~Mocz, Y.~Kahn, M.~Lisanti, and D.~N. Spergel, ``{Dynamical Friction in a Fuzzy Dark Matter Universe},'' \href{http://dx.doi.org/10.1088/1475-7516/2020/01/001}{{\em JCAP} {\bfseries 01} (2020) 001}, \href{http://arxiv.org/abs/1909.06381}{{\ttfamily arXiv:1909.06381 [astro-ph.CO]}}.

\bibitem{DuttaChowdhury:2023qxg}
D.~Dutta~Chowdhury, F.~C. van~den Bosch, P.~van Dokkum, V.~H. Robles, H.-Y. Schive, and T.~Chiueh, ``{On the Dynamical Heating of Dwarf Galaxies in a Fuzzy Dark Matter Halo},'' \href{http://dx.doi.org/10.3847/1538-4357/acc73d}{{\em Astrophys. J.} {\bfseries 949} no.~2, (2023) 68}, \href{http://arxiv.org/abs/2303.08846}{{\ttfamily arXiv:2303.08846 [astro-ph.GA]}}.

\bibitem{Zimmermann:2024xvd}
T.~Zimmermann, J.~Alvey, D.~J.~E. Marsh, M.~Fairbairn, and J.~I. Read, ``{Dwarf galaxies imply dark matter is heavier than $\mathbf{2.2 \times 10^{-21}} \, \mathbf{eV}$},'' \href{http://arxiv.org/abs/2405.20374}{{\ttfamily arXiv:2405.20374 [astro-ph.CO]}}.

\bibitem{Teodori:2025rul}
L.~Teodori, A.~Caputo, and K.~Blum, ``{Ultra-Light Dark Matter Simulations and Stellar Dynamics: Tension in Dwarf Galaxies for $m < 5\times10^{-21} $ eV},'' \href{http://arxiv.org/abs/2501.07631}{{\ttfamily arXiv:2501.07631 [astro-ph.GA]}}.

\bibitem{Dalal:2022rmp}
N.~Dalal and A.~Kravtsov, ``{Excluding fuzzy dark matter with sizes and stellar kinematics of ultrafaint dwarf galaxies},'' \href{http://dx.doi.org/10.1103/PhysRevD.106.063517}{{\em Phys. Rev. D} {\bfseries 106} no.~6, (2022) 063517}, \href{http://arxiv.org/abs/2203.05750}{{\ttfamily arXiv:2203.05750 [astro-ph.CO]}}.

\bibitem{Hui:2021tkt}
L.~Hui, ``{Wave Dark Matter},'' \href{http://dx.doi.org/10.1146/annurev-astro-120920-010024}{{\em Ann. Rev. Astron. Astrophys.} {\bfseries 59} (2021) 247--289}, \href{http://arxiv.org/abs/2101.11735}{{\ttfamily arXiv:2101.11735 [astro-ph.CO]}}.

\bibitem{Ferreira:2020fam}
E.~G.~M. Ferreira, ``{Ultra-light dark matter},'' \href{http://dx.doi.org/10.1007/s00159-021-00135-6}{{\em Astron. Astrophys. Rev.} {\bfseries 29} no.~1, (2021) 7}, \href{http://arxiv.org/abs/2005.03254}{{\ttfamily arXiv:2005.03254 [astro-ph.CO]}}.

\bibitem{Guzman2004}
F.~S. Guzman and L.~A. Urena-Lopez, ``{Evolution of the Schrodinger-Newton system for a selfgravitating scalar field},'' \href{http://dx.doi.org/10.1103/PhysRevD.69.124033}{{\em Phys. Rev.} {\bfseries D69} (2004) 124033},
\href{http://arxiv.org/abs/gr-qc/0404014}{{\ttfamily arXiv:gr-qc/0404014 [gr-qc]}}.

\bibitem{Schive:2014dra}
H.-Y. Schive, T.~Chiueh, and T.~Broadhurst, ``{Cosmic Structure as the Quantum Interference of a Coherent Dark Wave},'' \href{http://dx.doi.org/10.1038/nphys2996}{{\em Nature Phys.} {\bfseries 10} (2014) 496--499},
\href{http://arxiv.org/abs/1406.6586}{{\ttfamily arXiv:1406.6586 [astro-ph.GA]}}.

\bibitem{Schive:2014hza}
H.-Y. Schive, M.-H. Liao, T.-P. Woo, S.-K. Wong, T.~Chiueh, T.~Broadhurst, and W.~Y.~P. Hwang, ``{Understanding the Core-Halo Relation of Quantum Wave Dark Matter from 3D Simulations},'' \href{http://dx.doi.org/10.1103/PhysRevLett.113.261302}{{\em Phys. Rev. Lett.} {\bfseries 113} no.~26, (2014) 261302},
\href{http://arxiv.org/abs/1407.7762}{{\ttfamily arXiv:1407.7762 [astro-ph.GA]}}.

\bibitem{Schwabe:2016rze}
B.~Schwabe, J.~C. Niemeyer, and J.~F. Engels, ``{Simulations of solitonic core mergers in ultralight axion dark matter cosmologies},'' \href{http://dx.doi.org/10.1103/PhysRevD.94.043513}{{\em Phys. Rev.} {\bfseries D94} no.~4, (2016) 043513},
\href{http://arxiv.org/abs/1606.05151}{{\ttfamily arXiv:1606.05151 [astro-ph.CO]}}.

\bibitem{Veltmaat:2016rxo}
J.~Veltmaat and J.~C. Niemeyer, ``{Cosmological particle-in-cell simulations with ultralight axion dark matter},'' \href{http://dx.doi.org/10.1103/PhysRevD.94.123523}{{\em Phys. Rev.} {\bfseries D94} no.~12, (2016) 123523},
\href{http://arxiv.org/abs/1608.00802}{{\ttfamily arXiv:1608.00802 [astro-ph.CO]}}.

\bibitem{Mocz:2017wlg}
P.~Mocz, M.~Vogelsberger, V.~H. Robles, J.~Zavala, M.~Boylan-Kolchin, A.~Fialkov, and L.~Hernquist, ``{Galaxy formation with BECDM: I. Turbulence and relaxation of idealized haloes},'' \href{http://dx.doi.org/10.1093/mnras/stx1887}{{\em Mon. Not. Roy. Astron. Soc.} {\bfseries 471} no.~4, (2017) 4559--4570},
\href{http://arxiv.org/abs/1705.05845}{{\ttfamily arXiv:1705.05845 [astro-ph.CO]}}.

\bibitem{Veltmaat:2018dfz}
J.~Veltmaat, J.~C. Niemeyer, and B.~Schwabe, ``{Formation and structure of ultralight bosonic dark matter halos},'' \href{http://dx.doi.org/10.1103/PhysRevD.98.043509}{{\em Phys. Rev. D} {\bfseries 98} no.~4, (2018) 043509}, \href{http://arxiv.org/abs/1804.09647}{{\ttfamily arXiv:1804.09647 [astro-ph.CO]}}.

\bibitem{Levkov:2018kau}
D.~G. Levkov, A.~G. Panin, and I.~I. Tkachev, ``{Gravitational Bose-Einstein condensation in the kinetic regime},'' \href{http://dx.doi.org/10.1103/PhysRevLett.121.151301}{{\em Phys. Rev. Lett.} {\bfseries 121} no.~15, (2018) 151301},
\href{http://arxiv.org/abs/1804.05857}{{\ttfamily arXiv:1804.05857 [astro-ph.CO]}}.

\bibitem{Eggemeier:2019jsu}
B.~Eggemeier and J.~C. Niemeyer, ``{Formation and mass growth of axion stars in axion miniclusters},'' \href{http://dx.doi.org/10.1103/PhysRevD.100.063528}{{\em Phys. Rev. D} {\bfseries 100} no.~6, (2019) 063528}, \href{http://arxiv.org/abs/1906.01348}{{\ttfamily arXiv:1906.01348 [astro-ph.CO]}}.

\bibitem{Chen:2020cef}
J.~Chen, X.~Du, E.~W. Lentz, D.~J.~E. Marsh, and J.~C. Niemeyer, ``{New insights into the formation and growth of boson stars in dark matter halos},'' \href{http://dx.doi.org/10.1103/PhysRevD.104.083022}{{\em Phys. Rev. D} {\bfseries 104} no.~8, (2021) 083022}, \href{http://arxiv.org/abs/2011.01333}{{\ttfamily arXiv:2011.01333 [astro-ph.CO]}}.

\bibitem{Schwabe:2020eac}
B.~Schwabe, M.~Gosenca, C.~Behrens, J.~C. Niemeyer, and R.~Easther, ``{Simulating mixed fuzzy and cold dark matter},'' \href{http://dx.doi.org/10.1103/PhysRevD.102.083518}{{\em Phys. Rev. D} {\bfseries 102} no.~8, (2020) 083518}, \href{http://arxiv.org/abs/2007.08256}{{\ttfamily arXiv:2007.08256 [astro-ph.CO]}}.

\bibitem{Zhang:2018ghp}
J.~Zhang, H.~Liu, and M.-C. Chu, ``{Cosmological Simulation for Fuzzy Dark Matter Model},'' \href{http://dx.doi.org/10.3389/fspas.2018.00048}{{\em Front. Astron. Space Sci.} {\bfseries 5} (2019) 48}, \href{http://arxiv.org/abs/1809.09848}{{\ttfamily arXiv:1809.09848 [astro-ph.CO]}}.

\bibitem{Schwabe:2021jne}
B.~Schwabe and J.~C. Niemeyer, ``{Deep Zoom-In Simulation of a Fuzzy Dark Matter Galactic Halo},'' \href{http://dx.doi.org/10.1103/PhysRevLett.128.181301}{{\em Phys. Rev. Lett.} {\bfseries 128} no.~18, (2022) 181301}, \href{http://arxiv.org/abs/2110.09145}{{\ttfamily arXiv:2110.09145 [astro-ph.CO]}}.

\bibitem{Bar:2018acw}
N.~Bar, D.~Blas, K.~Blum, and S.~Sibiryakov, ``{Galactic rotation curves versus ultralight dark matter: Implications of the soliton-host halo relation},'' \href{http://dx.doi.org/10.1103/PhysRevD.98.083027}{{\em Phys. Rev.} {\bfseries D98} no.~8, (2018) 083027},
\href{http://arxiv.org/abs/1805.00122}{{\ttfamily arXiv:1805.00122 [astro-ph.CO]}}.

\bibitem{Bar:2021kti}
N.~Bar, K.~Blum, and C.~Sun, ``{Galactic rotation curves versus ultralight dark matter: A systematic comparison with SPARC data},'' \href{http://dx.doi.org/10.1103/PhysRevD.105.083015}{{\em Phys. Rev. D} {\bfseries 105} no.~8, (2022) 083015}, \href{http://arxiv.org/abs/2111.03070}{{\ttfamily arXiv:2111.03070 [hep-ph]}}.

\bibitem{Banares-Hernandez:2023axy}
A.~Ba\~nares Hern\'andez, A.~Castillo, J.~Martin~Camalich, and G.~Iorio, ``{Confronting fuzzy dark matter with the rotation curves of nearby dwarf irregular galaxies},'' \href{http://dx.doi.org/10.1051/0004-6361/202346686}{{\em Astron. Astrophys.} {\bfseries 676} (2023) A63}, \href{http://arxiv.org/abs/2304.05793}{{\ttfamily arXiv:2304.05793 [astro-ph.GA]}}.

\bibitem{Marsh:2015wka}
D.~J.~E. Marsh and A.-R. Pop, ``{Axion dark matter, solitons and the cusp-core problem},'' \href{http://dx.doi.org/10.1093/mnras/stv1050}{{\em Mon. Not. Roy. Astron. Soc.} {\bfseries 451} no.~3, (2015) 2479--2492},
\href{http://arxiv.org/abs/1502.03456}{{\ttfamily arXiv:1502.03456 [astro-ph.CO]}}.

\bibitem{Blum:2021oxj}
K.~Blum and L.~Teodori, ``{Gravitational lensing H0 tension from ultralight axion galactic cores},'' \href{http://dx.doi.org/10.1103/PhysRevD.104.123011}{{\em Phys. Rev. D} {\bfseries 104} no.~12, (2021) 123011}, \href{http://arxiv.org/abs/2105.10873}{{\ttfamily arXiv:2105.10873 [astro-ph.CO]}}.

\bibitem{Blum:2024igb}
K.~Blum and L.~Teodori, ``{Hunting for ultralight dark matter with cosmographic H0 signal},'' \href{http://dx.doi.org/10.1103/PhysRevD.111.043509}{{\em Phys. Rev. D} {\bfseries 111} no.~4, (2025) 043509}, \href{http://arxiv.org/abs/2409.04134}{{\ttfamily arXiv:2409.04134 [astro-ph.CO]}}.

\bibitem{Li:2020ryg}
X.~Li, L.~Hui, and T.~D. Yavetz, ``{Oscillations and Random Walk of the Soliton Core in a Fuzzy Dark Matter Halo},'' \href{http://dx.doi.org/10.1103/PhysRevD.103.023508}{{\em Phys. Rev. D} {\bfseries 103} no.~2, (2021) 023508}, \href{http://arxiv.org/abs/2011.11416}{{\ttfamily arXiv:2011.11416 [astro-ph.CO]}}.

\bibitem{Chowdhury:2021zik}
D.~D. Chowdhury, F.~C. van~den Bosch, V.~H. Robles, P.~van Dokkum, H.-Y. Schive, T.~Chiueh, and T.~Broadhurst, ``{On the Random Motion of Nuclear Objects in a Fuzzy Dark Matter Halo},'' \href{http://dx.doi.org/10.3847/1538-4357/ac043f}{{\em Astrophys. J.} {\bfseries 916} no.~1, (2021) 27}, \href{http://arxiv.org/abs/2105.05268}{{\ttfamily arXiv:2105.05268 [astro-ph.GA]}}.

\bibitem{Chan:2023crj}
J.~H.-H. Chan, S.~Sibiryakov, and W.~Xue, ``{Boson star normal modes},'' \href{http://dx.doi.org/10.1007/JHEP08(2023)045}{{\em JHEP} {\bfseries 08} (2023) 045}, \href{http://arxiv.org/abs/2304.13054}{{\ttfamily arXiv:2304.13054 [astro-ph.CO]}}.

\bibitem{Khmelnitsky:2013lxt}
A.~Khmelnitsky and V.~Rubakov, ``{Pulsar timing signal from ultralight scalar dark matter},'' \href{http://dx.doi.org/10.1088/1475-7516/2014/02/019}{{\em JCAP} {\bfseries 1402} (2014) 019},
\href{http://arxiv.org/abs/1309.5888}{{\ttfamily arXiv:1309.5888 [astro-ph.CO]}}.

\bibitem{Schutz:2020jox}
K.~Schutz, ``{Subhalo mass function and ultralight bosonic dark matter},'' \href{http://dx.doi.org/10.1103/PhysRevD.101.123026}{{\em Phys. Rev. D} {\bfseries 101} no.~12, (2020) 123026}, \href{http://arxiv.org/abs/2001.05503}{{\ttfamily arXiv:2001.05503 [astro-ph.CO]}}.

\bibitem{Laroche:2022pjm}
A.~Laroche, D.~Gilman, X.~Li, J.~Bovy, and X.~Du, ``{Quantum fluctuations masquerade as halos: Bounds on ultra-light dark matter from quadruply-imaged quasars},'' \href{http://arxiv.org/abs/2206.11269}{{\ttfamily arXiv:2206.11269 [astro-ph.CO]}}.

\bibitem{Powell:2023jns}
D.~M. Powell, S.~Vegetti, J.~P. McKean, S.~D.~M. White, E.~G.~M. Ferreira, S.~May, and C.~Spingola, ``{A lensed radio jet at milli-arcsecond resolution II: Constraints on fuzzy dark matter from an extended gravitational arc},'' \href{http://arxiv.org/abs/2302.10941}{{\ttfamily arXiv:2302.10941 [astro-ph.CO]}}.

\bibitem{Blas:2024duy}
D.~Blas, S.~Gasparotto, and R.~Vicente, ``{Searching for ultra-light dark matter through frequency modulation of gravitational waves},'' \href{http://arxiv.org/abs/2410.07330}{{\ttfamily arXiv:2410.07330 [hep-ph]}}.

\bibitem{Hertzberg:2022vhk}
M.~P. Hertzberg and A.~Loeb, ``{Quantum tunneling of ultralight dark matter out of satellite galaxies},'' \href{http://dx.doi.org/10.1088/1475-7516/2023/02/059}{{\em JCAP} {\bfseries 02} (2023) 059}, \href{http://arxiv.org/abs/2212.07386}{{\ttfamily arXiv:2212.07386 [astro-ph.CO]}}.

\bibitem{Kendall:2019fep}
E.~Kendall and R.~Easther, ``{The Core-Cusp Problem Revisited: ULDM vs. CDM},'' \href{http://dx.doi.org/10.1017/pasa.2020.3}{{\em Publ. Astron. Soc. Austral.} {\bfseries 37} (2020) e009}, \href{http://arxiv.org/abs/1908.02508}{{\ttfamily arXiv:1908.02508 [astro-ph.CO]}}.

\bibitem{Navarro:1996gj}
J.~F. Navarro, C.~S. Frenk, and S.~D.~M. White, ``{A Universal density profile from hierarchical clustering},'' \href{http://dx.doi.org/10.1086/304888}{{\em Astrophys. J.} {\bfseries 490} (1997) 493--508},
\href{http://arxiv.org/abs/astro-ph/9611107}{{\ttfamily arXiv:astro-ph/9611107 [astro-ph]}}.

\bibitem{Moore:1994yx}
B.~Moore, ``{Evidence against dissipationless dark matter from observations of galaxy haloes},'' \href{http://dx.doi.org/10.1038/370629a0}{{\em Nature} {\bfseries 370} (1994) 629}.

\bibitem{Bar-Or:2018pxz}
B.~Bar-Or, J.-B. Fouvry, and S.~Tremaine, ``{Relaxation in a Fuzzy Dark Matter Halo},'' \href{http://dx.doi.org/10.3847/1538-4357/aaf28c}{{\em Astrophys. J.} {\bfseries 871} (2019) 28},
\href{http://arxiv.org/abs/1809.07673}{{\ttfamily arXiv:1809.07673 [astro-ph.GA]}}.

\bibitem{Bar-Or:2020tys}
B.~Bar-Or, J.-B. Fouvry, and S.~Tremaine, ``{Relaxation in a Fuzzy Dark Matter Halo. II. Self-consistent Kinetic Equations},'' \href{http://dx.doi.org/10.3847/1538-4357/abfb66}{{\em Astrophys. J.} {\bfseries 915} no.~1, (2021) 27}, \href{http://arxiv.org/abs/2010.10212}{{\ttfamily arXiv:2010.10212 [astro-ph.GA]}}.

\bibitem{Dmitriev:2023ipv}
A.~S. Dmitriev, D.~G. Levkov, A.~G. Panin, and I.~I. Tkachev, ``{Self-Similar Growth of Bose Stars},'' \href{http://dx.doi.org/10.1103/PhysRevLett.132.091001}{{\em Phys. Rev. Lett.} {\bfseries 132} no.~9, (2024) 091001}, \href{http://arxiv.org/abs/2305.01005}{{\ttfamily arXiv:2305.01005 [astro-ph.CO]}}.

\bibitem{Chavanis:2020upb}
P.-H. Chavanis, ``{Landau equation for self-gravitating classical and quantum particles: application to dark matter},'' \href{http://dx.doi.org/10.1140/epjp/s13360-021-01617-3}{{\em Eur. Phys. J. Plus} {\bfseries 136} no.~6, (2021) 703}, \href{http://arxiv.org/abs/2012.12858}{{\ttfamily arXiv:2012.12858 [astro-ph.GA]}}.

\bibitem{May:2021wwp}
S.~May and V.~Springel, ``{Structure formation in large-volume cosmological simulations of fuzzy dark matter: impact of the non-linear dynamics},'' \href{http://dx.doi.org/10.1093/mnras/stab1764}{{\em Mon. Not. Roy. Astron. Soc.} {\bfseries 506} no.~2, (2021) 2603--2618}, \href{http://arxiv.org/abs/2101.01828}{{\ttfamily arXiv:2101.01828 [astro-ph.CO]}}.

\bibitem{Veltmaat:2019hou}
J.~Veltmaat, B.~Schwabe, and J.~C. Niemeyer, ``{Baryon-driven growth of solitonic cores in fuzzy dark matter halos},'' \href{http://dx.doi.org/10.1103/PhysRevD.101.083518}{{\em Phys. Rev. D} {\bfseries 101} no.~8, (2020) 083518}, \href{http://arxiv.org/abs/1911.09614}{{\ttfamily arXiv:1911.09614 [astro-ph.CO]}}.

\bibitem{Liao:2024zkj}
P.-Y. Liao, G.-M. Su, H.-Y. Schive, A.~Kunkel, H.~Huang, and T.~Chiueh, ``{Deciphering the Soliton-Halo Relation in Fuzzy Dark Matter},'' \href{http://arxiv.org/abs/2412.09908}{{\ttfamily arXiv:2412.09908 [astro-ph.CO]}}.

\bibitem{Du:2016aik}
X.~Du, C.~Behrens, J.~C. Niemeyer, and B.~Schwabe, ``{Core-halo mass relation of ultralight axion dark matter from merger history},'' \href{http://dx.doi.org/10.1103/PhysRevD.95.043519}{{\em Phys. Rev.} {\bfseries D95} no.~4, (2017) 043519},
\href{http://arxiv.org/abs/1609.09414}{{\ttfamily arXiv:1609.09414 [astro-ph.GA]}}.

\bibitem{Chan:2021bja}
H.~Y.~J. Chan, E.~G.~M. Ferreira, S.~May, K.~Hayashi, and M.~Chiba, ``{The diversity of core\textendash{}halo structure in the fuzzy dark matter model},'' \href{http://dx.doi.org/10.1093/mnras/stac063}{{\em Mon. Not. Roy. Astron. Soc.} {\bfseries 511} no.~1, (2022) 943--952}, \href{http://arxiv.org/abs/2110.11882}{{\ttfamily arXiv:2110.11882 [astro-ph.CO]}}.

\bibitem{Zagorac:2022xic}
J.~L. Zagorac, E.~Kendall, N.~Padmanabhan, and R.~Easther, ``{Soliton formation and the core-halo mass relation: An eigenstate perspective},'' \href{http://dx.doi.org/10.1103/PhysRevD.107.083513}{{\em Phys. Rev. D} {\bfseries 107} no.~8, (2023) 083513}, \href{http://arxiv.org/abs/2212.09349}{{\ttfamily arXiv:2212.09349 [astro-ph.CO]}}.

\bibitem{Lague:2023wes}
A.~Lagu\"e, B.~Schwabe, R.~Hlo\v{z}ek, D.~J.~E. Marsh, and K.~K. Rogers, ``{Cosmological simulations of mixed ultralight dark matter},'' \href{http://dx.doi.org/10.1103/PhysRevD.109.043507}{{\em Phys. Rev. D} {\bfseries 109} no.~4, (2024) 043507}, \href{http://arxiv.org/abs/2310.20000}{{\ttfamily arXiv:2310.20000 [astro-ph.CO]}}.

\bibitem{Manita:2024vww}
Y.~Manita, T.~Takahashi, and A.~Taruya, ``{Soliton self-gravity and core-halo relation in fuzzy dark matter halos},'' \href{http://arxiv.org/abs/2411.14614}{{\ttfamily arXiv:2411.14614 [astro-ph.CO]}}.

\bibitem{Yavetz:2021pbc}
T.~D. Yavetz, X.~Li, and L.~Hui, ``{Construction of wave dark matter halos: Numerical algorithm and analytical constraints},'' \href{http://dx.doi.org/10.1103/PhysRevD.105.023512}{{\em Phys. Rev. D} {\bfseries 105} no.~2, (2022) 023512}, \href{http://arxiv.org/abs/2109.06125}{{\ttfamily arXiv:2109.06125 [astro-ph.CO]}}.

\bibitem{Chavanis:2019faf}
P.-H. Chavanis, ``{Derivation of the core mass -- halo mass relation of fermionic and bosonic dark matter halos from an effective thermodynamical model},'' \href{http://dx.doi.org/10.1103/PhysRevD.100.123506}{{\em Phys. Rev. D} {\bfseries 100} no.~12, (2019) 123506}, \href{http://arxiv.org/abs/1905.08137}{{\ttfamily arXiv:1905.08137 [astro-ph.CO]}}.

\bibitem{Chan:2022bkz}
J.~H.-H. Chan, S.~Sibiryakov, and W.~Xue, ``{Condensation and evaporation of boson stars},'' \href{http://dx.doi.org/10.1007/JHEP01(2024)071}{{\em JHEP} {\bfseries 01} (2024) 071}, \href{http://arxiv.org/abs/2207.04057}{{\ttfamily arXiv:2207.04057 [astro-ph.CO]}}.

\bibitem{Salucci:2000ps}
P.~Salucci and A.~Burkert, ``{Dark matter scaling relations},'' \href{http://dx.doi.org/10.1086/312747}{{\em Astrophys. J. Lett.} {\bfseries 537} (2000) L9--L12}, \href{http://arxiv.org/abs/astro-ph/0004397}{{\ttfamily arXiv:astro-ph/0004397}}.

\bibitem{Burkert:2015vla}
A.~Burkert, ``{The Structure and Dark Halo Core Properties of Dwarf Spheroidal Galaxies},'' \href{http://dx.doi.org/10.1088/0004-637X/808/2/158}{{\em Astrophys. J.} {\bfseries 808} no.~2, (2015) 158}, \href{http://arxiv.org/abs/1501.06604}{{\ttfamily arXiv:1501.06604 [astro-ph.GA]}}.

\bibitem{Walker:2008ax}
M.~G. Walker, M.~Mateo, and E.~Olszewski, ``{Stellar Velocities in the Carina, Fornax, Sculptor and Sextans dSph Galaxies: Data from the Magellan/MMFS Survey},'' \href{http://dx.doi.org/10.1088/0004-6256/137/2/3100}{{\em Astron. J.} {\bfseries 137} (2009) 3100}, \href{http://arxiv.org/abs/0811.0118}{{\ttfamily arXiv:0811.0118 [astro-ph]}}.

\bibitem{Walker:2009zp}
M.~G. Walker, M.~Mateo, E.~W. Olszewski, J.~Penarrubia, N.~W. Evans, and G.~Gilmore, ``{A Universal Mass Profile for Dwarf Spheroidal Galaxies},'' \href{http://dx.doi.org/10.1088/0004-637X/704/2/1274}{{\em Astrophys. J.} {\bfseries 704} (2009) 1274--1287}, \href{http://arxiv.org/abs/0906.0341}{{\ttfamily arXiv:0906.0341 [astro-ph.CO]}}. [Erratum: Astrophys.J. 710, 886--890 (2010)].

\bibitem{Bar:2019bqz}
N.~Bar, K.~Blum, J.~Eby, and R.~Sato, ``{Ultra-light dark matter in disk galaxies},''
\href{http://arxiv.org/abs/1903.03402}{{\ttfamily arXiv:1903.03402 [astro-ph.CO]}}.

\bibitem{bryan1998statistical}
G.~L. Bryan and M.~L. Norman, ``Statistical properties of x-ray clusters: Analytic and numerical comparisons,'' {\em The Astrophysical Journal} {\bfseries 495} no.~1, (1998) 80.

\bibitem{Nori:2020jzx}
M.~Nori and M.~Baldi, ``{Scaling relations of fuzzy dark matter haloes \textendash{} I. Individual systems in their cosmological environment},'' \href{http://dx.doi.org/10.1093/mnras/staa3772}{{\em Mon. Not. Roy. Astron. Soc.} {\bfseries 501} no.~1, (2021) 1539--1556}, \href{http://arxiv.org/abs/2007.01316}{{\ttfamily arXiv:2007.01316 [astro-ph.CO]}}.

\bibitem{Mina:2020eik}
M.~Mina, D.~F. Mota, and H.~A. Winther, ``{Solitons in the dark: First approach to non-linear structure formation with fuzzy dark matter},'' \href{http://dx.doi.org/10.1051/0004-6361/202038876}{{\em Astron. Astrophys.} {\bfseries 662} (2022) A29}, \href{http://arxiv.org/abs/2007.04119}{{\ttfamily arXiv:2007.04119 [astro-ph.CO]}}.

\bibitem{Mocz:2019pyf}
P.~Mocz {\em et~al.}, ``{First star-forming structures in fuzzy cosmic filaments},'' \href{http://dx.doi.org/10.1103/PhysRevLett.123.141301}{{\em Phys. Rev. Lett.} {\bfseries 123} no.~14, (2019) 141301}, \href{http://arxiv.org/abs/1910.01653}{{\ttfamily arXiv:1910.01653 [astro-ph.GA]}}.

\bibitem{Kolb:1993zz}
E.~W. Kolb and I.~I. Tkachev, ``{Axion miniclusters and Bose stars},'' \href{http://dx.doi.org/10.1103/PhysRevLett.71.3051}{{\em Phys. Rev. Lett.} {\bfseries 71} (1993) 3051--3054}, \href{http://arxiv.org/abs/hep-ph/9303313}{{\ttfamily arXiv:hep-ph/9303313}}.

\bibitem{Gorghetto:2024vnp}
M.~Gorghetto, E.~Hardy, and G.~Villadoro, ``{More axion stars from strings},'' \href{http://dx.doi.org/10.1007/JHEP08(2024)126}{{\em JHEP} {\bfseries 08} (2024) 126}, \href{http://arxiv.org/abs/2405.19389}{{\ttfamily arXiv:2405.19389 [hep-ph]}}.

\bibitem{Gorghetto:2022sue}
M.~Gorghetto, E.~Hardy, J.~March-Russell, N.~Song, and S.~M. West, ``{Dark photon stars: formation and role as dark matter substructure},'' \href{http://dx.doi.org/10.1088/1475-7516/2022/08/018}{{\em JCAP} {\bfseries 08} no.~08, (2022) 018}, \href{http://arxiv.org/abs/2203.10100}{{\ttfamily arXiv:2203.10100 [hep-ph]}}.

\bibitem{Amin:2022pzv}
M.~A. Amin, M.~Jain, R.~Karur, and P.~Mocz, ``{Small-scale structure in vector dark matter},'' \href{http://dx.doi.org/10.1088/1475-7516/2022/08/014}{{\em JCAP} {\bfseries 08} no.~08, (2022) 014}, \href{http://arxiv.org/abs/2203.11935}{{\ttfamily arXiv:2203.11935 [astro-ph.CO]}}.

\bibitem{10.1145/301631.301661}
M.~Frigo, ``A fast fourier transform compiler,'' \href{http://dx.doi.org/10.1145/301631.301661}{{\em SIGPLAN Not.} {\bfseries 34} no.~5, (May, 1999) 169–180}. \url{https://doi.org/10.1145/301631.301661}.

\bibitem{Widrow:1993qq}
L.~M. Widrow and N.~Kaiser, ``{Using the Schrodinger equation to simulate collisionless matter},'' {\em Astrophys. J. Lett.} {\bfseries 416} (1993) L71--L74.

\end{thebibliography}\endgroup
\bibliographystyle{utphys}

\end{document}